\documentclass[traditabstract]{aa}

\usepackage{array}
\usepackage{graphicx}
\usepackage{txfonts}
\usepackage{natbib}
\bibpunct{(}{)}{;}{a}{}{,} 

\def\Offset{\mathcal{O}}
\def\SecondQuant{S_\mathrm{q}}

\def\Qerrdiff{\epsilon_{q,\mathrm{diff}}}

\begin{document}

\title{Planck-LFI radiometers tuning}
\titlerunning{~LFI Radiometers -- Tuning}

\author{
F.~Cuttaia\inst{1}
\and A.~Mennella\inst{2}
\and L.~Stringhetti\inst{1}
\and M.~Maris\inst{3}
\and L.~Terenzi\inst{1}
\and M.~Tomasi\inst{2}
\and F.~Villa\inst{1}
\and M.~Bersanelli\inst{2}
\and R.C.~Butler\inst{1}
\and B.~Cappellini\inst{2}
\and L.P~Cuevas\inst{10}
\and O.~D'Arcangelo\inst{4}
\and R.~Davis\inst{7}
\and M.~Frailis\inst{3}
\and C.~Franceschet\inst{2}
\and E.~Franceschi\inst{1}
\and A.~Gregorio\inst{5} 
\and R.~Hoyland\inst{2} 
\and R.~Leonardi\inst{6} 
\and S.~Lowe\inst{7}
\and N.~Mandolesi\inst{1}
\and P.~Meinhold\inst{6}
\and L.~Mendes\inst{9}
\and and N.~Roddis\inst{7}
\and M.~Sandri\inst{1}
\and L.~Valenziano\inst{1}
\and A.~Wilkinson\inst{7}
\and A.~Zacchei\inst{3}
\and A.~Zonca\inst{2}
\and P.~Battaglia\inst{8}
\and S.~De Nardo\inst{8}
\and S.~Grassi\inst{8}
\and M.~Lapolla\inst{8}
\and P.~Leutenegger\inst{8}
\and M.~Miccolis\inst{8}
\and R.~Silvestri\inst{8}
}
 \authorrunning{F.~Cuttaia et al.}

\institute{
  Istituto di Astrofisica Spaziale e Fisica Cosmica, INAF, via P.~Gobetti 101 -- I-40129~Bologna, Italy\\
  \email{cuttaia@iasfbo.inaf.it}
\and 
  Universit\`{a} degli Studi di Milano, via Celoria 16 -- I-20133~Milano, Italy  
\and
  INAF~/~OATS, via Tiepolo 11 -- I-34143~Trieste, Italy
\and
  IFP-CNR, via Cozzi 53 -- I-20013~Milano, Italy
\and
  University of Trieste, Department of Physics, via Valerio 2 -- I-34127~Trieste, Italy
\and
  Department of Physics, University of California, Santa Barbara, CA~93106-9530, USA
\and
  Jodrell Bank Centre for Astrophysics, Alan Turing Building, The University of Manchester, Manchester, M13~9PL, UK  
\and
  Thales Alenia Space Italia S.p.A., IUEL - Scientific Instruments, S.S. Padana Superiore 290 -- I-20090~Vimodrone (MI), Italy
\and
  Planck Science Office, European Space Agency, ESAC, P.O.~box~78, 28691 Villanueva de la Ca\~{n}ada, Madrid, Spain  
\and
  Research and Scientific Support Department of ESA, ESTEC, Noordwijk, The Netherlands  }

\date{Received DD MMMM 2008 / Accepted DD MMMM 2009}

\abstract{
This paper describes the Planck Low Frequency Instrument tuning
activities performed through the ground test campaigns,
from Unit to Satellite Levels. Tuning is key to achieve the best
possible instrument performance and tuning parameters strongly depend on
thermal and electrical conditions. For this reason tuning has been
repeated several times during ground tests and it has been repeated in
flight before starting nominal operations. The paper discusses the
tuning philosophy, the activities and the obtained results, highlighting
developments and changes occurred during test campaigns. The paper
concludes with an overview of tuning performed during the satellite
cryogenic test campaign (Summer 2008) and of the plans for the just
started in-flight calibration.}


\maketitle
%

\section{Introduction}

    PLANCK represents the third generation of mm-wave instruments designed for space observations of CMB anisotropies within the new Cosmic Vision 2020 ESA Science Programme. PLANCK was successfully launched on 2009 May the 14th, carrying the state of the art of microwave radiometers (Low Frequency Instrument, \citep[see][]{2009_LFI_cal_M1}) and bolometers (High Frequency Instrument,\citep[see][]{2009_HFI_PAPER}) operating between 30 GHz and 900 GHz, in nine frequency channels, coupled with a 1.5 m telescope \citep[see][]{2009_Tauber_Planck_Optics})\\
   	The Low Frequency Instrument is a system of 22 wide band radio receivers, based on Indium Phosphide HEMTs amplifiers, adopting a pseudo-correlation scheme reducing 1/f noise due to gain fluctuations \citep[see][]{Seiffert02} , covering the range 30 GHz - 70 GHz.\\
    Maximum instrument performance can be achieved after proper tuning, that consists in finding the best front-end biases and the optimal parameters controlling the back-end electronics. Because these parameters depend both on the input signal characteristics and on the instrument thermal and electrical boundary conditions, tuning has been carried out at various integration levels and will be also repeated in flight before starting nominal operations.\\
    Tuning is achieved by performing several steps in series: first the best front-end bias voltages and currents are found, then the parameters controlling signal digitization in the Data Acquisition Electronics (DAE) box and finally the digital compression parameters are optimised. In this paper we discuss this process in detail, presenting the main results obtained from the instrument and satellite test campaigns. In particular we show how the tuning philosophy has improved during the test campaign and provide an overview of the final strategy conceived for the flight Calibration and Performance Verification (CPV).

\section{LFI architecture and Tuning}
\label{ARCH-setup}

    \subsection{Instrument references}
    \label{INST-REF}

        Because the subject is treated with a high level of technical details, in Table~\ref{tab:list_reference_papers} we provide a list of the main reference papers.

        \begin{table}[h!]
            \caption{List of the main reference papers}
            \label{tab:list_reference_papers}
            \begin{center}
                \begin{tabular}{l l}
                    \hline
                    \hline
                    Topic & Reference \\
                    \hline
                    LFI instrument description &\cite {2009_LFI_cal_M2}\\
                    Instrument calibration and performance &\cite{2009_LFI_cal_M3}\\
                    30-44~GHz Front ends &\cite{2009_LFI_cal_R8}\\
                    30-44 GHz back ends &\cite{2009_LFI_cal_R9}\\
                    70 GHz receiver &\cite{2009_LFI_cal_R10}\\
                    Back-end digital electronics (REBA) &\cite{2009_LFI_cal_REBA}\\
                    Cryo-facility for receiver-level tests &\cite{2009_LFI_cal_T1}\\
                    Cryo-facility for instrument-level tests &\cite{2009_LFI_cal_T2}\\
                    Sky-load simulator for receiver-level tests &\cite{2009_LFI_cal_T4}\\
                    Sky-load simulator for instrument-level tests &\cite{2009_LFI_cal_T5}\\
                    Instrument operation software &\cite{2009_LFI_cal_DE}\\
                    Data analysis software &\cite{2009_LFI_cal_D4}\\
                    Digital compression optimisation &\cite{2009_LFI_cal_D2}\\
                    \hline
                \end{tabular}
            \end{center}
        \end{table}

    \subsection{Instrument Description}
    \label{INST-DESCR}
    
        From the point of view of instrument tuning, the LFI \citep[see][]{2009_LFI_cal_M2} can be subdivided in three main parts: the receiver array consisting of 22 pseudo correlation differential radiometers, the electronics box for analog signal conditioning and digitisation (DAE) and the digital signal processing unit (REBA). Each receiver collects the sky radiation through a corrugated dual profiled feed-horn \citep[see][]{2009_LFI_cal_O1} connected to an ortho-mode transducer \citep[see][]{2009_LFI_cal_O2} that separates the two orthogonal polarisations which propagate through two independent radiometers, each one terminating with two detector diodes.

\begin{figure*}
            \begin{center}
                \resizebox{\hsize}{!}{\includegraphics{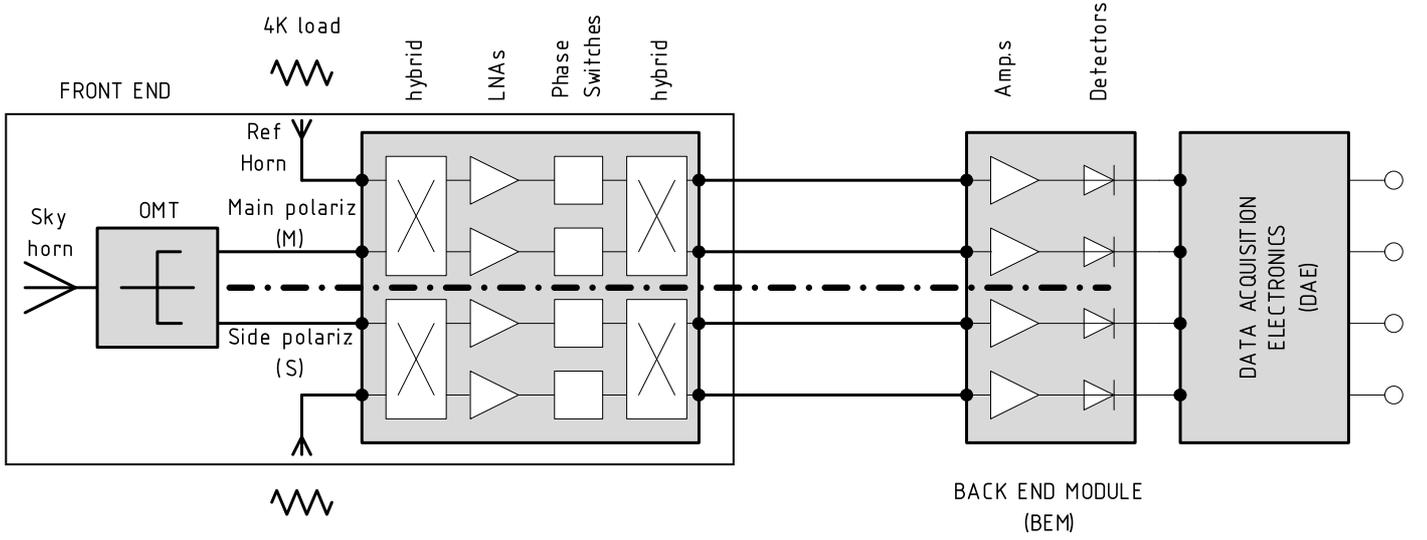}}
                \caption{Scheme of one LFI Channel. In the black rectangle is represented the Front End, cooled at about 20K, containing, from left to right: the sky feed horn, the Ortho Mode Transducer separating the sky signal in two orthogonal polarizations entering two radiometers having the same design (the dot dashed line separates them) Each radiometer contains: one reference horn looking one reference load at 4K, two hybrids connected through two LNAs amplifiers and two 180 deg phase switches. Outside the black rectangle is the warm (about 300K) Back End, containing the Back End Modules (filtering, amplifying and integrating the analog signals received from the FEMs through stainless steel waveguides, thermally decoupling the cold and warm units) and the DAE (further amplifying the analog signal and converting it into digital before sending to the REBA).
}
                \label{rca_scheme}
            \end{center}
        \end{figure*}

        In each radiometer, the sky signal is coupled with that from a stable reference load at $\sim$4~K \citep[see][]{2009_LFI_cal_R1} and amplified by $\sim$70~dB with InP HEMT\footnote{Indium Phosphide High Electron Mobility Transistor} amplifiers located in a Front-end unit cooled to $\sim$20~K thanks to a closed cycle Hydrogen sorption cooler (see \cite{2009_LFI_cal_T2}, \cite{2004_Sorcio}) and in a warm ($\sim$300~K) Back-end unit. Back-end bias voltages are fixed, while front-end biases (transistor voltages and phase switch diode currents) can be controlled and optimised to obtain maximum noise performance. A detailed block diagram is presented in Fig.~\ref{rca_scheme}.

\begin{figure}[h!]
            \begin{center}
                \includegraphics[width = 8cm]{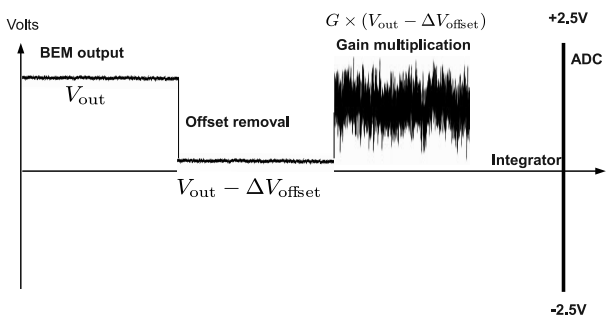}
                \caption{Schematic of the post-detection analog signal conditioning}
                \label{fig:post_detection_analog_signal_conditioning}
            \end{center}
        \end{figure}

        The 44 detector diodes are connected to the DAE that has the main function to digitise the analog signals into 14-bit integers. Before digitization, a dedicated circuit removes a constant offset and applies a gain factor to each detector output in order to otimise the signal to the analog to digital converter (ADC) dynamic range (see Fig.~\ref{fig:post_detection_analog_signal_conditioning}). The offset and gain values are two programmable parameters that must be tuned for each channel in order to obtain an average voltage output slightly above zero and optimal noise resolution ($\sim$75\% of the ADC dynamic range).
        
        After digitization, data are further quantised and compressed before building telemetry packets in order to be compliant with the available bandwidth. This step is critical for the scientific performance and a specific test has been performed to find the optimal parameters. 

        Finally, in this paper we will often refer to receivers and individual channels according to a nomenclature that is described in Appendix~\ref{app:naming_convention}.

    \subsection{Tuning flow}
    \label{TUN-FLOW}
    
        The various parameters needing optimisation must be tuned in series, starting from front-end biases and ending with the digital compression parameters. This flow is represented schematically in Table~\ref{tab:tuning_flow}.

        \begin{table}[h!]
            \caption{List of instrument tuning parameters and optimisation sequence}
            \label{tab:tuning_flow}
            \begin{tabular}{l p{3cm} l}
                \hline
                \textbf{Front-end biases} & Phase switch bias currents & $I_1$, $I_2$\\
                                          & Amplifier (drain and gate) bias voltages & $V_{\rm d}$, $V_{\rm g1}$, $V_{\rm g2}$\\
                \textbf{Analog signal processing} & Programmable gain and offset & $V_0$, $G_{\rm DAE}$\\
                \textbf{Digital signal processing} & Signal mixing parameters & $r_1$, $r_2$\\
                                                   & Digital quantisation parameters & $\SecondQuant$, $\Offset$\\
                \hline
            \end{tabular}
        \end{table}

        Because optimal parameters depend critically on the input signal characteristics and on thermal and electrical boundary conditions, tuning activities have been performed several times during the integration and test campaign on individual sub-units (amplifiers, Front End and Back End modules), on each individual Radiometer Chain Assembly (RCA) and on the integrated Radiometer Array Assembly (RAA).\\
        
        Tuning at Unit level and at RCA level was performed in different times and locations, depending on the groups responsible of the development: 30/44 GHz Front End modules were developed in UK- Manchester, 30/44 GHz Back-End modules in Spain - Santander, 70 GHz Front End and Back End modules in Finland - Helsinki. Tests on 30/44 GHz channels and 70 GHz channels followed two parallel ways, either for the Qualification Models ( QM) and for the Flight Models (FM).\\
        
         The 30/44 GHz FM Front End Units were integrated with Back End Units and tested together for the first time only in 2005 during the Test campaign at Radiometer Chain Level (RCA) in Alcatel Alenia Space - Laben , Milan. During RCA level tests a DAE Breadboard was used and no compression was applied to the digitized data. After that, 30/44/70 GHz channels were integrated together and tested at Instrument Level (RAA) together with DAE and REBA Flight Model, nominal and redundant units.\\
        
        Once integrated on the Planck satellite, the instrument has been further tuned during the satellite cryogenic test campaign, integrated with the HFI into the Planck payload, (test campaign performed at the Centre Spatiale de Li\`ege during Summer 2008) and , at the time of this paper, is being tuned for the last time in flight during the ongoing Calibration and Performance Verification (CPV) phase, before starting nominal operations.\\


    The tuning strategy has been continuously improved during the long test campaign \citep[see][]{2009_LFI_cal_M3} and adapted to the cryogenic setups used at the various integration levels. Major changes accounted for the different design of the cryo-chambers as, for instance, the temperature of cryo-environment, of the reference loads and of the sky simulators, the design of the thermal controllers, the electric cross talk in the bias power suppliers, etc. (For more details see \cite{2009_LFI_cal_R8}, \cite{2009_LFI_cal_T1} and \cite{2009_LFI_cal_T2}).\\   
     
        In this paper we will present in detail tuning activities, with particular emphasis on the instrument-level test campaign carried out at the Thales Alenia Space Italia laboratories located in Vimodrone (Milano) during 2006, and during the last on-ground tests at satellite level performed at the Centre Spatial de Liege in Summer 2008.


\section{Front-end bias tuning}
\label{FEM-TUNING}

    Front-end biases are the first parameters that are optimised in order to obtain the best possible receiver sensitivity and isolation\footnote{Isolation is the receiver ability to separate the sky and reference load signals downstream of the second hybrid coupler.}. Sensitivity is mainly determined by the noise contribution of the first amplification stage in the front-end amplifier assembly, while isolation mostly depends on the gain balance in the radiometer legs between the two hybrid couplers (refer to Fig.~\ref{rca_scheme}).

    The first step is finding, for each phase switch, the bias currents ($I_1$, $I_2$) to the two diodes that provide minimal and balanced insertion losses in the two switch states in order to minimise any contribution from the phase switch imperfect isolation.

\subsection{Phase switch  tuning}
\label{PH-TUN}

    \subsubsection{Theory}
    \label{PHSW-TUN-THEORY}
    
\begin{figure}[h!]
            \begin{center}
                \resizebox{\hsize}{!}{\includegraphics{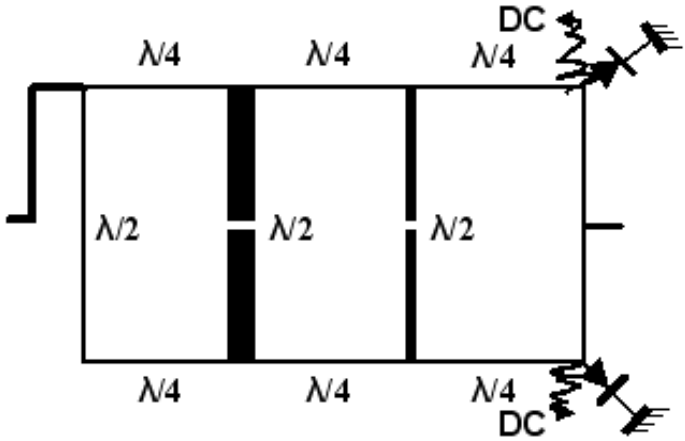}}
                \caption{Phase switch scheme. The design is the same for all the channels and is based on InP monolithic
microwave integrated circuit (MMIC) chip version manufactured on the HBT InP wafer process at NGST. 
The basic design consists in two interconnected hybrid rings, providing a $180^{\circ}$ phase switch difference with phase error $\pm 1^{\circ}$, very good insertion loss ($\leq 2.5 dB$) , return loss $\le -10dB$) and bandwidth ( $\ge 20\%$ of the nominal wavelength ). The use of the two hybrids enhances the matching between the input port ( accepting radiation  from the 1st Hybrid amplified in the HEMT LNAs) and the output port (connected to the 2nd Hybrid). Ports in the 1st millimetric circuit and in the 2nd  symmetric circuit are connected through transmission lines, whose length and width are fractions of the wavelengths of the specific band and multiple of the characteristic impedance of the system they belong to. The phase switch design is under Patent US 6,803,838 B2 - Oct. 12,2004
}
                \label{fig_PHSW}
            \end{center}
        \end{figure}

The Phase switches \citep[see][]{2003_hoyland_espoo}), between the LNA and the second hybrid, use pin diodes (the scheme in Fig.~\ref{fig_PHSW}) allowing radiation to travel directly or  lambda/2 longer (introducing 180 deg phase shift), depending on their polarization. They are controlled by changing the two currents $I_1$, $I_2$ (controlling the amount of RF power flowing through diodes) and the state of polarization (1 direct, 0 inverse). \

        The amplitude match in the two switch states is achieved by switching on only one FEM amplifier at a time, when the sky and reference signals are no longer separated by the second hybrid so that the signals in the two phase switch states (named here as ``even'' and ``odd'' states) at each of the hybrid outputs can be written as \citep[see][]{Seiffert02}): 
        
         \begin{eqnarray}
            S_{a(b)}^{\rm even} \propto \sqrt{A_2} e^{i\phi_2} G_{\rm FEM} \left ( N_{\rm FEM}  + {S_{\rm sky} + S_{\rm ref} \over \sqrt{2} }\right ) \label{eq:psw_tuning_output1} \\
            S_{a(b)}^{\rm odd} \propto  \sqrt{A_1} e^{i\phi_1} G_{\rm FEM} \left ( N_{\rm FEM}  + {S_{\rm sky} + S_{\rm ref} \over \sqrt{2} }\right ),
            \label{eq:psw_tuning_output2}
        \end{eqnarray}
        where $a$ and $b$ indicate the two detectors connected to the two hybrid outputs.
                
        Considering that the two phase switch states correspond to $\phi_1 = 0^\circ$ and $\phi_2 = 180^\circ$, from Eq.~(\ref{eq:psw_tuning_output1}) and ~(\ref{eq:psw_tuning_output2}) we see that the difference in power $\delta = \left | S_{a(b)}^{\rm even} \right |^2 - \left | S_{a(b)}^{\rm odd} \right |^2$ is zero at each of the two radiometer detector diodes if $A_1 = A_2$, i.e. when the phase switch is balanced.

        The phase switch can therefore be balanced by finding the pair of diode currents, $I_1$ and $I_2$, that minimise the quadratic sum:

        \begin{equation}
            \delta_{\rm r.m.s.} = \sqrt {\delta_a^2 + \delta_b^2}
            \label{eq:psw_tuning_delta_rms}
        \end{equation}        
        Because the phase switch are tuned before the front-end amplifiers, we are clearly assuming that the amplitude matching is independent of the bias applied to the low noise amplifiers. Although we believe that possible interactions do not affect tuning to first order, we will verify this assumption in flight, during which the phase switch tuning will be run both before and after the front-end amplifiers Tuning.

        \begin{figure}[h!]
            \begin{center}
                \resizebox{\hsize}{!}{\includegraphics{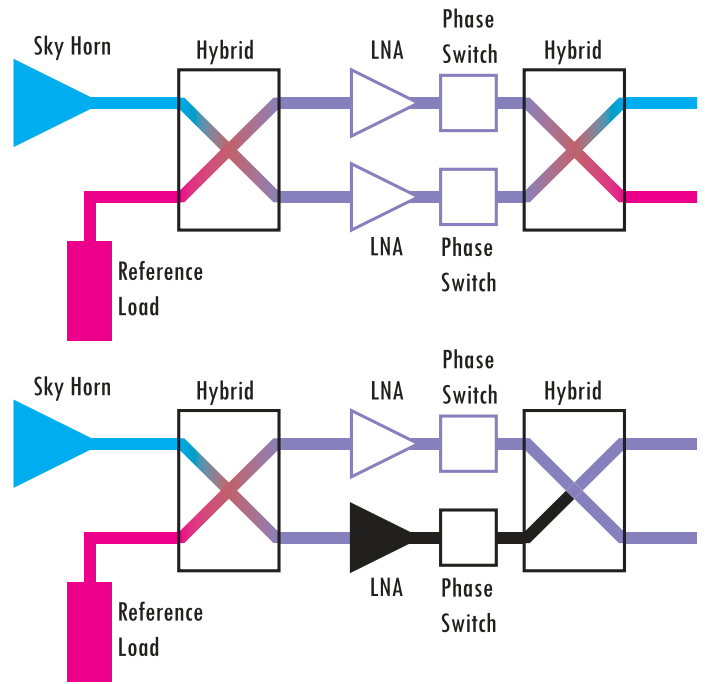}}
                \caption{
                    Phase switch tuning schematic. In the upper figure the radiometer is working in its nominal conditions so that the sky and reference signals are disentangled by  the $2^{\rm nd}$ hybrid. The figure represents a snapshot in a particular switch state, while a snapshot of the next state would have the red and blue signals inverted. The bottom figure shows the condition in which one amplifier is switched off. Under this condition,  the signals at the output of both hybrid arms remain a combination of both sky and reference signals (changes in  the matching between the phase switch and the LNA off, possibly producing spurious differences in the output when the two polarizations are exercised, are negligible because of the low unwanted cross talk in the hybrid, less than -15 dB). Therefore no separation in different phase switch states should be observed, unless the phase switches is not well balanced in amplitude.
                }
                \label{fig3-c}
            \end{center}
        \end{figure}

    \subsubsection{Results }
    \label{PHSW-TUN-RES}
        
        In this section we report results of phase switch tuning performed on 30 and 44~GHz receivers on single receivers \citep[see][]{2009_LFI_cal_M4}) , on the integrated instrument and during tests conducted at satellite level. Phase switches of 70~GHz radiometers \citep[see][]{2009_LFI_cal_R10}) , in fact, were not tuned and their diodes were set at the maximum bias currents. This choice was driven by the switch time response that in the 70~GHz devices was longer compared to the 30 and 44~GHz and further increased when the diode currents were lowered.

        Because a long phase switch transient typically results in a shorter integration time, we preferred to minimize such transients and accept a small imbalance in the phase switch amplitudes (which can be corrected for, at first order, by a proper amplifier gain balance).
               
        A critical point in the tuning procedure is represented by the choice of the sampling strategy of the $[I_1,\, I_2]$ bidimensional bias space that consists of 255$\times$255 equally spaced bias combinations between 0 and $\sim$1~mA. Different strategies were chosen during the various test phases in order to optimise the width of the sampled region and the time available for the test. 

        For example, during the test performed on individual receivers, the experimental setup only allowed manual control of phase switch biases so that full sampling of the bias space was not feasible. In that case the solution was to change $I_1$ and $I_2$ in opposite directions around the starting value provided by the manufacturer. A similar strategy was also followed during the instrument-level tests.\\      
        Results, however, showed that optimal solution are not unique, i.e. multiple pairs $[I_1,\, I_2]$ can minimise $\delta_{\rm r.m.s.}$ in Eq.~(\ref{eq:psw_tuning_delta_rms}). 
        
        This highlighted the importance of increasing the sampled region in order to fully characterise the region of optimal biases and prompted a substantial improvement of the tuning procedure by implementing a square matrix sampling during satellite-level tests, on ground and in flight. Differences can have multiple causes, such as the modified sampling strategy of the bias currents or the possible bias shifts related with the thermal distribution along the cryo harness powering the Front End. This argument will apply again in the next paragraph when describing the LNAs Tuning.

\begin{figure}[h!]
            \begin{center}
                \resizebox{\hsize}{!}{\includegraphics{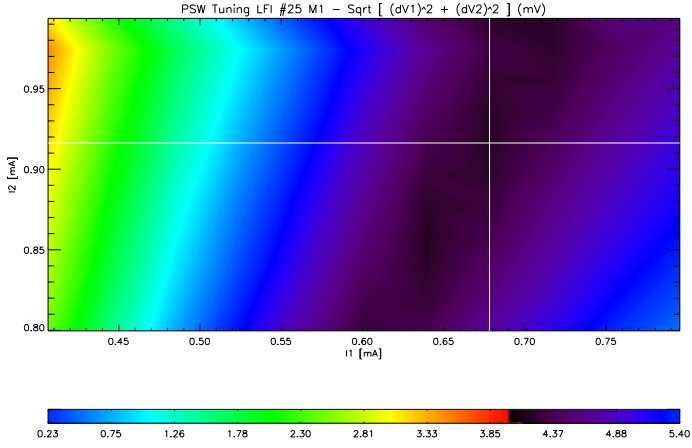}}
                \caption{Example of results from phase switch tuning performed during satellite tests. The optimal balancing is obtained for $I_1$=0.678 mA and $I_2$=0.916 mA}
                \label{fig6-1}
            \end{center}        
        \end{figure}

        In Fig.~\ref{fig6-1} we show a colour plot of $\delta_{\rm r.m.s.}$ as a function of $I_1$ and $I_2$ sampled in a plane. The dark ``valley'' represents the region of minimum $\delta_{\rm r.m.s.}$, i.e. where the phase switch amplitudes are balanced. The actual bias currents have then been chosen in order to minimise the phase switch insertion loss compatible with the power consumption constraints. 
        
\begin{figure}[h!]
            \begin{center}
                \resizebox{\hsize}{!}{\includegraphics{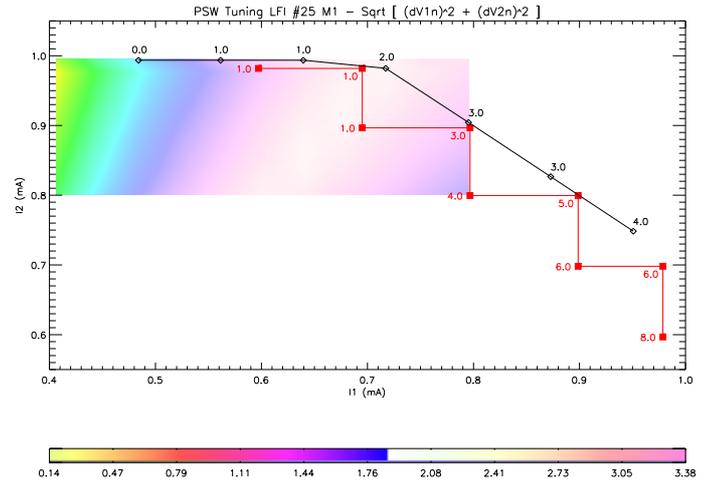}}
                \caption{Comparison between phase switch tuning results performed during tests at different levels. The coloured region shows the range scanned during matrix phase switch Tuning at satellite level. Red filled squares represent the sampling performed at receiver level test (RCA), while black empty diamonds are relative to Instrument level phase switch tuning (RAA). The labels on the RCA and RAA points, as the CSL coloured contour, measure the phase switch unbalancing as: $\delta_{\rm r.m.s.}^* = \sqrt  {\frac{\delta(V_1)^2}{V_1^2} + \frac{\delta (V_2^2}{V_2^2}}\times100$. The bias region containing optimal $I_1$,$I_2$ pairs do not coincide for the three data sets, confirming the need to retune the phase switches at any time the setup conditions are changed.}
                \label{fig7-1}                          
            \end{center}
        \end{figure}

        In Fig.~\ref{fig7-1} we show a comparison of the phase switch tuning samplings during on ground receiver-level, instrument-level and satellite-level tests.\\
        
        Full results are reported in Appendix~\ref{app:EX-SUP} (see Tab.~\ref{tab:PHSW_Tuning_comparison})

    \subsection{Amplifiers bias tuning}
    \label{LNA-TUN}
    
\begin{figure}[h!]
            \begin{center}
                \includegraphics[width=7.5cm]{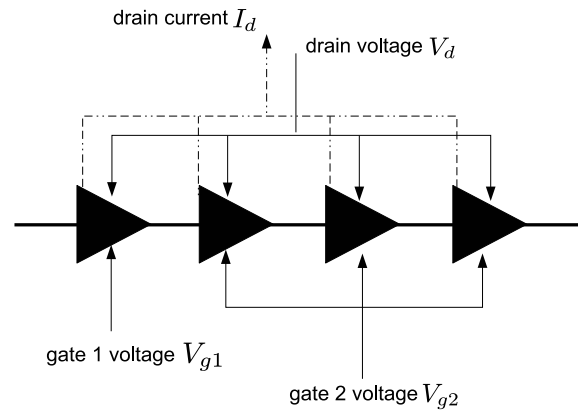}
            \end{center}
            \caption{
                Schematic of an ACA (Amplifier Chain Assembly. Each assembly is composed by four (two in 70 GHz and five in 44 GHz channels) amplifier stages driven by a common drain voltage, a gate voltage to the first stage and a common gate voltage to the other stages. The total drain current flowing in the ACA is measured and provided in the housekeeping data: actually, this is the only housekeeping measured at FEM level, since bias voltages are measured only at DAE drivers level.
            }
            \label{fig:schematic_aca}
        \end{figure}

        Each Amplifier Chain Assembly (ACA) is composed of four low noise InP amplifiers and is driven by three voltages: a common drain voltage ($V_{\rm d}$), a gate voltage for the first stage ($V_{\rm g1}$) and a common gate voltage for the remaining stages      ($V_{\rm g2}$) (see Fig.~\ref{fig:schematic_aca}). The total drain current $I_{\rm d}$ flowing in the ACA is measured and is available in the instrument housekeeping. 
        
        The bias voltages to the various stages have been independently tuned during the ACA assembly, before wiring together the independent stages. Hence, once the optimal biases were determined at unit level for each amplifier, their ratio has been definitely fixed by the potentiometry circuit. Therefore, all subsequent tuning activities have aimed at optimising voltage biases with respect to the particular thermal and electrical environmental changes among the different test setups.

        In particular, because several ACAs share a common bias return line, we find that bias changes on one amplifier can affect also the others. Furthermore the bias voltage readout is done at the DAE box and not at the front-end terminals: this implies that the actual bias reaching the FEM can be estimated only by knowing the voltage drop along the harness lines. A model was specifically developed in order to predict such voltage drops, but the complexity of the thermal environment and of the grounding scheme strongly limited its accuracy. Therefore amplifier tuning has been a mandatory step any time the instrument was tested in a different thermal or electrical environment.

        Because $V_{\rm d}$ is the bias most impacting the power consumption, it was tuned during tests at device level and has not been changed; noise and gain balance optimisations have been performed by adjusting $V_{\rm g1}$ and $V_{\rm g2}$ by measuring the receiver noise temperature and isolation for a number of bias voltage combinations around the optimal point found during the RCA test campaign. 

\begin{figure}[h!]
            \begin{center}
                \resizebox{\hsize}{!}{\includegraphics{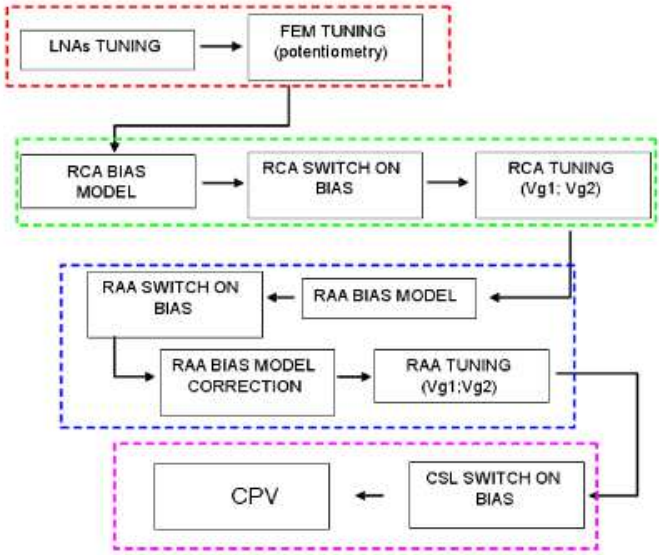}}
                \caption{
                    Amplifier bias tuning flow from unit level tests to receiver and instrument level tests, which provided the optimal biases used for switching the instrument on during satellite tests. Dashed coloured boxes indicate the different phases. Unit level tuning, from 2004 to 2006, red; RCA level tuning, from 2005 to 2006, green; Instrument level tuning, 2006, blue; satellite level tuning, performed on-ground during Summer 2008 and ongoing in flight during CPV, magenta.
                }   
                  \label{tuning_phases}     
            \end{center}
                 \end{figure}

        The overall tuning activities carried out during all the LFI integration and test campaign is shortly summarized in  Fig.~\ref{tuning_phases}. Each tuning phase assumes, as a starting point, the optimal bias points obtained in the previous level and a model that allows prediction at first order of the voltages at the Front-end module starting from the bias at the level of the DAE box.

    \subsubsection{Basic Theory}
    \label{LNA-THEO}
    
        During receiver-level and instrument-level tests amplifiers have been tuned in two steps, the first aimed at finding, for each amplifier, the $V_{\rm g1}$ optimising noise figure, and the second to find the values of $V_{\rm g2}$ in the two amplifier legs that maximise gain balance and, therefore, receiver isolation. The basic assumption behind this strategy is that the first stage essentially determines the amplifier noise, while the remaining stages mostly contribute to the gain, with minimal effects on noise.

        \paragraph{Noise temperature.} Noise temperature of a $N$-stage amplifier can be defined as follows:
        
        \begin{equation}
            T_n = T_{n1}+\frac{T_{n2}}{G_1} +\frac{T_{n3}}{G_1\,G_2}+\ldots
            +\frac{T_{nN}}{G_1\,G_2\,\ldots\,G_{N-1}},
            \label{eq:noise_temperature_amplifier}
        \end{equation}
        where $T_{n1}\ldots T_{nN}$ and $G_1\ldots G_N$ are the noise temperatures and gains of each of the cascaded amplifiers. 

        The noise temperature can be measured by the well known $Y$-factor method, based upon the voltage output recorded at two different thermal inputs of either the sky or the reference load:
%
        \begin{equation}
            T_n=\frac{T_{\rm high} - Y\times T_{\rm low}}{Y-1},
            \label{eq:y_factor}
        \end{equation}
        where $Y = V_{\rm high} / V_{\rm low}$

        \paragraph{Isolation.} Isolation represents a measure of the ability of the pseudo-correlator to separate the sky and reference load signals after the second hybrid. In fact, if the hybrid phase matching is not perfect and/or the gains of the two radiometer arms are not balanced, then the separation after the second hybrid is not perfect and a certain level of mixing between the two signals will be present in the output \citep[see][]{mennella03}).
        
        Isolation can be determined experimentally by changing the temperature of one of the two loads and measuring any variation induced in the signal nominally coming from the stable load. If we perform the test by changing the sky load temperature, isolation is given by:
        
        \begin{equation}
            I\approx \frac{\Delta V_{\rm ref}}{\Delta V_{\rm sky} + \Delta V_{\rm ref}}.
            \label{eq:isolation_simple}
        \end{equation}
        
        In case also the stable temperature load experiences spurious variations (e.g. given by non perfect thermal decoupling between the sky and reference loads) we can correct Eq.~(\ref{eq:isolation_simple}) if we know the receiver photometric constant $G_0$ (in the limit of linear response):
  
        \begin{equation}
            I\approx \frac{\Delta V_{\rm ref} - G_0\, \Delta T_{\rm ref}}{\Delta V_{\rm sky}+\Delta V_{\rm ref}-G_0\, \Delta T_{\rm ref}},
            \label{eq:isolation}
        \end{equation}
        which is valid if the temperature change $\Delta T_{\rm ref}$ is in a range where the radiometric response is linear.

\begin{figure}
            \begin{center}
                \resizebox{\hsize}{!}{\includegraphics{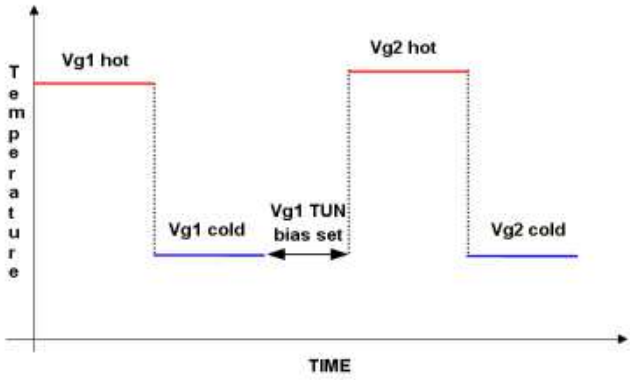}}
                \caption{
                    Experimental tuning sequence followed during radiometer and instrument level tests. Red and blue colours indicate hot and cold temperature states. Vg1 tuning data are processed and optimal bias were set before starting the Vg2 bias scan, in order to have the bias condition the closer to the final setting. Cooldown and warmup phases are represented by the black dot lines.
                }
                \label{fig9-1}
            \end{center}
        \end{figure}

        \paragraph{Experimental.} In Fig.~\ref{fig9-1} we show the experimental sequence adopted up to instrument-level test for tuning front-end amplifiers. During the Calibrations at RCA level and at Instrument level , two sky simulators were used to provide a stable thermal input to the feedhorns (see \cite{2009_LFI_cal_T4}, \cite {2009_LFI_cal_T5}, \cite {2005_Phd_Cuttaia}).\\ 
        The sequence consists of two steps: each step corresponds to a temperature change (from hot to cold or in the opposite direction) of one of the loads (depending on the cryogenic setup, sky load or the reference load was used). 
          In the first step we tune $V_{\rm g1}$ for each amplifier: the same set of bias values is run in each of the two temperature states. During this test, the ACA paired with the one under test is kept off, similarly to what is done during the phase switch tuning procedure (see Section~\ref{PHSW-TUN-THEORY}). In this case the receiver does not separate sky and reference signals and the radiometer voltage output is proportional to the average of the sky and reference load temperatures. If, for example, we perform the test at two different temperatures of the sky load, $T_{\rm sky,1}$ and $T_{\rm sky,2}$ with $T_{\rm sky,1} > T_{\rm sky,2}$, in Eq.~(\ref{eq:y_factor}) we have that $T_{\rm high}= \frac{T_{\rm sky,1}+T_{\rm ref}}{2}$ and $T_{\rm low}= \frac{T_{\rm sky,2}+T_{\rm ref}}{2}$.
          For each bias point, the Noise Temperature is calculated and is used as figure of merit to choose the optimal $V_{\rm g1}$ setting for each radiometer.
        
        After the optimal $V_{\rm g1}$ values are found and set in the instrument, a second temperature step is performed in order to find the optimal $V_{\rm g2}$. During each temperature step, the receiver is operated nominally and data are acquired for a set of $V_{\rm g2}$ values. For each value Isolation is computed according to Eq.~(\ref{eq:isolation}) and, for each radiometer, the bias pair corresponding to the minimum is chosen. Fig.~\ref{fig10-c} shows the scheme followed.

    \subsubsection{Results from instrument-level tests}
    \label{VG1}
    
        \paragraph{Gate 1 voltage.} Tuning of the gate 1 voltage has been performed by exploiting a temperature jump in the sky load of $\sim 8$~K, from $\sim 21$~K to $\sim 29$~K. The back-end and reference load temperatures ranged in the intervals $[37.5 - 38.0]$~C and $[22.1 - 22.2]$~K, respectively, while the front-end unit was at 26.4~K with a stability of $\pm 5$~mK.

\begin{figure}
  \resizebox{\hsize}{!}{\includegraphics{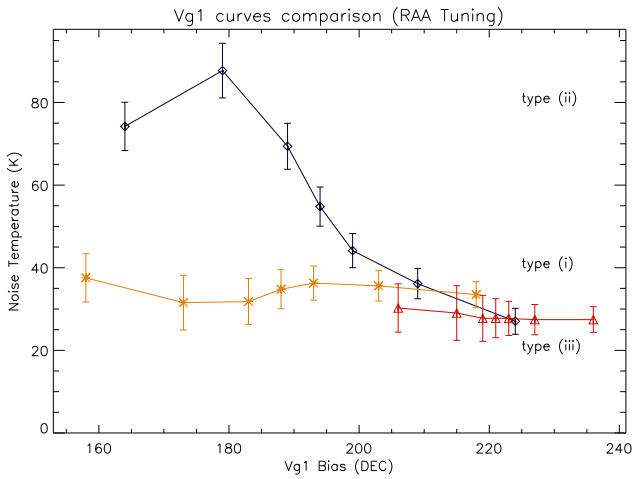}}
  \caption{Comparison between Vg1 curve shapes: Yellow curve, type (i) RCA 22M2, black curve, type (ii)  RCA 20 S1 ,red curve, type (iii) RCA 25. Error bars account for the uncertainty in the temperature of the changing load.}
  \label{fig13-c}
\end{figure}


        It must be stressed that a clear minimum in noise temperature was not always found in the data. Noise temperature curves showed sometimes a rather flat response or a monotone behaviour in the region where the minimum noise temperature is measured,      suggesting a possible absolute minimum outside the explored range, mostly in the direction of higher voltages (see Fig.~\ref{fig13-c}). In such cases a decision was taken according to the following guidelines:
    
    \begin{itemize}
        \item minimum noise temperature in the tested range;
        \item drain current close to the value expected from the receiver-level tests;
        \item optimal gain balance (verified by comparing drain currents and voltage outputs).
    \end{itemize}
    
\begin{table}[h!]
            \caption{Summary of Vg1 tuning behaviours grouped per channels. For simplicity, curves are grouped in three schematic categories: the behaviour refers to the regions showing the best noise temperatures.}
            \label{tab:Vg1_behavior}
            \begin{center}
                \begin{tabular}{l l}
                    \hline
                    \hline
                    Curve Type & Channels\\
                    \hline
                    (i)~defined minimum & RCA21 M1,M2 RCA22 M1, M2\\
                     ~                 & RCA23 S1,M2, RCA24, RCA\\
                     ~                 & 25M1, S2, RCA27 S1 \\
                    (ii)~roughly monotons       & RCA18 S1,S1, RCA19, RCA20\\
                     ~                 & RCA21 M2,RCA22 S2,RCA23 M1\\
                     ~                 & RCA26 S1,RCA27 S2,RCA28 S1, S2\\
                    (iii)~flat / wide minimum & RCA21 S2,RCA22 S1\\
                     ~                 & RCA25 M1,M2,S1,.RCA26 M1,M2,S2\\
                     ~                 & RCA27 M1,M2,.RCA28 M1,M2 \\
                                     
                    \hline
                \end{tabular}
            \end{center}
        \end{table}

\begin{figure}
         \includegraphics[width = 8cm]{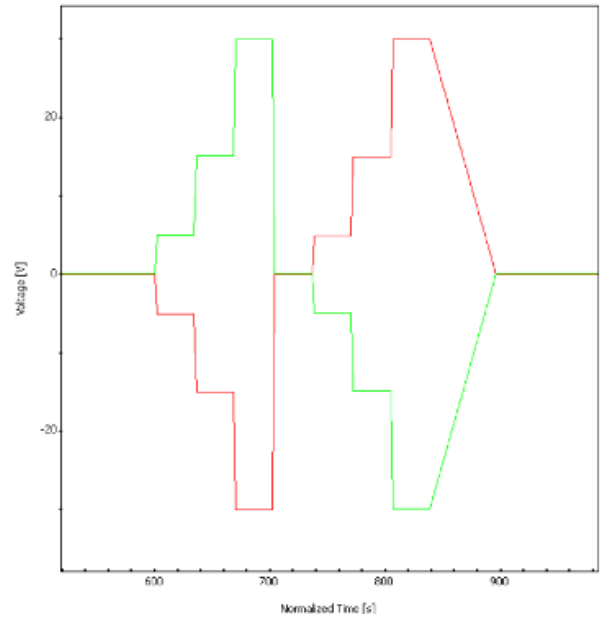}
         \caption{Isolation Tuning scheme followed at RCA and RAA test level: RCA 21 is represented here. Firstly Vg2(M1)- green-is increased and Vg2(M2)-red -decreased; hence the two quantities are varied in the opposite direction}
         \label{fig10-c}
 \end{figure}

        \paragraph{Gate 2 voltage.} Tuning of $V_{\rm g2}$ has been performed by exploiting a temperature jump in the sky load of $\sim 12$~K, from $\sim 18$~K to $\sim 30$~K, with back-end and reference load temperatures conditions similar to the $V_{\rm g1}$ tuning. The scheme followed, the same for each radiometer, is shown in Fig.~\ref{fig10-c}: Tuning was performed over one radiometer at a time, acting on the two paired chains. Biases were changed in opposite directions on the two ACAs, to prevent large variations in the total power budget (that would enhance drops in the bias lines)

        \subsubsection{Bias tuning at satellite level}
        
            Although the bias tuning strategy described in the previous sections proved successful during receiver and instrument level tests, it nevertheless showed to be unfeasible for satellite tests both on ground and in flight because, in these cases, a controlled temperature stage providing the necessary input steps is not available. 

            On the other hand, the cooldown profile of the HFI 4~K Stirling cooler \citep[see][]{4K_cooler}, which happens once during the satellite cooldown, provides a $\sim 16$~K single input temperature step at the reference loads between two stable states at $\sim 20$~K and $\sim 4$~K. This step can be exploited to measure noise temperature and isolation and, therefore, tune front-end biases. 

            The limited control available on the cooldown phase required, however, a deep revision of the entire tuning strategy that has been modified according to the following lines:

            \begin{itemize}
                \item $V_{\rm g1}$ and $V_{\rm g2}$ are not tuned in sequence but at the same time. This means that at each of the two stable temperature states each radiometer is operated in switching conditions with $V_{\rm g1}$ and $V_{\rm g2}$ changing on both amplifiers;
                
                \item after the temperature step is completed noise temperature and isolation are calculated for each bias combination and the one providing optimal performance is selected.
            \end{itemize}
            
\begin{figure}[h!]
                \begin{center}
                    \resizebox{\hsize}{!}{\includegraphics{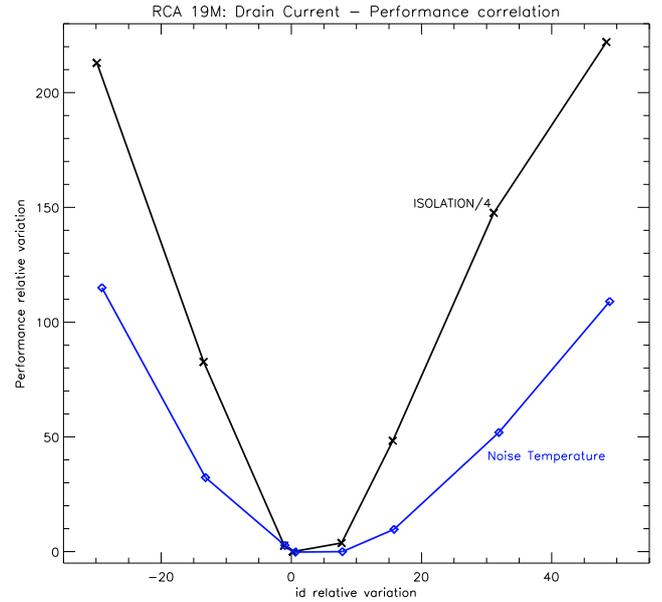}}\\
                    \caption{
                        Drain current - performance correlation from instrument-level data. Noise Temperature and Isolation variation  	around the optimal point is shown in \% on the y axis, drain current variation is shown in \% on the x axis. The correlation drain current - Noise Temperature is measured when varying $Vg_1$ while correlation drain current - Isolation while varying $Vg_2$. Since Isolation has a more evident dependance on Id, the value represented is one fourth of the true value, to make the comparison with Noise Temperature clearer.
                    }
                 \label{fig19-c}
                 \end{center}
            \end{figure}

            This approach, called \textit{Hyper Matrix Tuning}, is in principle very simple but carries the disadvantage that the number of potential bias combination for each radiometer scales with $N^4$ (where $N$ is the number of steps implemented for each of the two biases, $V_{\rm g1}$ and $V_{\rm g2}$) with potential impacts on the test schedule. Several optimisations have been implemented to fit the procedure into the available time:
            
            \begin{itemize}
                \item the integration time for each bias step is set at the minimum time (20'' for 70~GHz channels and 10'' for 30 and 44~GHz channels) necessary to avoid transient effects (which have been further minimised by properly sorting the bias steps according to minimum drain current changes);

                \item the bias sweep at the two stable temperature states is run in parallel on groups of receivers showing negligible mutual interaction: the electric susceptibility matrix was drawn, measuring the ratio of the induced voltage R = (induced voltage change)/ (voltage with tuned bias). The cut off value $\|$$R_0$$\|$ was chosen $\|$$R_0$$\|\le 1/20$ , that is the average voltage change produced by one bias step change on the same channel. Interactions were studied at the level of:\\
                
\begin{description}
	\item[ a-] Main and side arm of the same RCA.
	\item[ b-] RCAs belonging to the same power group.
	\item[ c-] RCAs belonging to different power groups but to the same FEM tray 		\citep[see][]{2009_LFI_cal_M2}. 
	\item[ d-] RCAs belonging to different power groups and different FEM trays.
\end{description}
         
         The grouping scheme adopted (d) did not show measurable electric coupling; no relevant interaction were observed also in (c), while large electric crosstalk ($\|R\|\geq 1/10$ in (a) and in the most cases of (b).\\
      
                \item Tuning data acquired during tests at receiver and instrument level have been used to reduce the bias space by excluding combinations providing extremely poor performance. In particular our tests showed that drain current is a good performance estimator which allowed us to exclude bias combinations by setting a threshold on the maximum deviation of $I_d$ from the design value. The correlation between $I_d$ and basic performance is shown in~Fig.\ref{fig19-c} for a representative case. 
            \end{itemize}

            A test campaign performed on spare radiometers showed that the same optimal biases were recovered with both strategies (i.e. with two temperature steps and with a single temperature jump): this strategy was successfully applied during ground cryogenic satellite tests and planned for flight calibration tests.
            
            \paragraph{Results from satellite level tests}
            During satellite-level tests the cooldown phase of the HFI 4~K cooler was used to run a simplified version of the Hyper Matrix Tuning. According to this approach, called \textit{Matrix Tuning} and chosen for schedule constraints, Vg1 and Vg2 biases are scanned for each radiometer first on the first amplifier and then on the second (and not simultaneously on both amplifiers). The clear advantage is that the test time scales with $2\, N^2$, but the main limitation is that a large fraction of the bias space is not tested thus increasing the risk of finding a local rather than the global performance maximum. 
            
\begin{figure}[h]
                \begin{center}
                    \resizebox{\hsize}{!}{\includegraphics{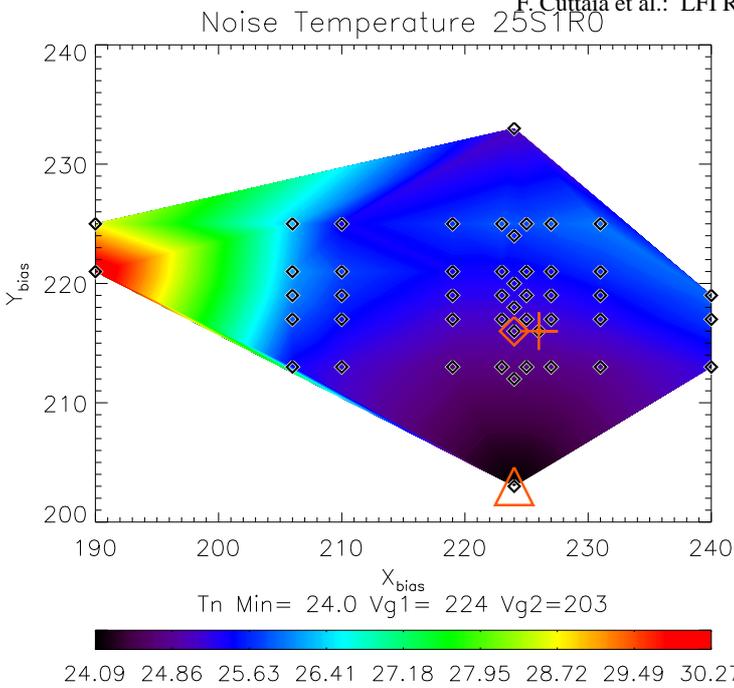}}
                    \caption{
                        Matrix tuning result relative to LFI25S1. $V_{\rm g1}$ and $V_{\rm g2}$ are plotted in decimal uncalibrated units on the x and y axes, while noise temperature is represented in colour scale. All the bias points explored are marked by small black diamonds. The optimal point found during instrument level tests is indicated in the subtitle and highlighted on the contour by a red triangle. Not always the bias point producing the lowest noise temperature was chosen: sometimes minima have been believed to be naive features of the procedure or unstable points. The optimal bias pair from RAA Tuning is marked by a large red cross while the bias pair chosen in CSL Tuning by a large red diamond.
                    }
                    \label{fig51-c}
                \end{center}            
     \end{figure}

            Four bias runs were performed at four different temperatures of the 4~K loads, so that the linearity response could be characterised. An example of results obtained during satellite tests is shown in Fig.~\ref{fig51-c}.

\begin{figure}[h!]
                \begin{center}
                    \resizebox{\hsize}{!}{\includegraphics{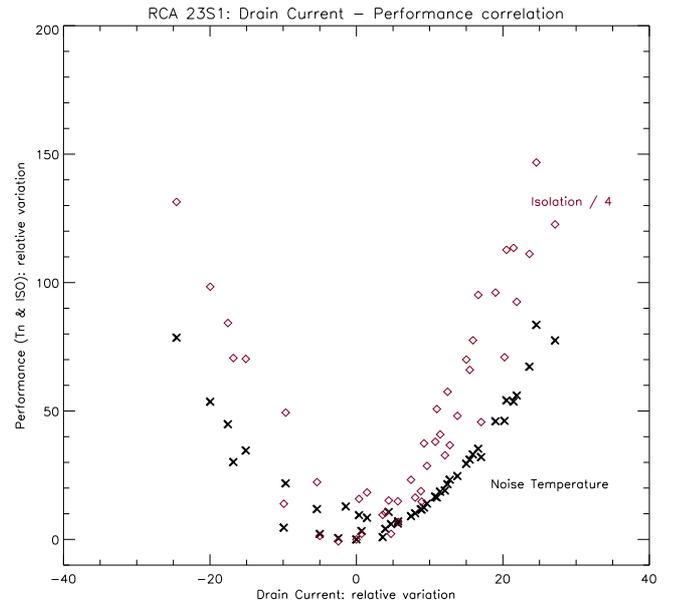}}
                    \caption{
                        LNAs drain current vs. performance correlation from satellite-level data. On the y axis the receiver performance (Noise temperature, in black crosses, and Isolation, in purple diamonds) is expressed in percent of the optimal value; on the x axis, drain current variation is expressed in percent of the value corresponding to best noise temperature. Since Isolation has a more evident dependence on Id, the value represented is one fourth of the true value, to make the comparison with Noise Temperature clearer. 
                    }
                 \label{fig19-c1}
                 \end{center}
            \end{figure}

Also for the matrix Tuning, correlation between Drain current and basic performance is very strong. However, in this case, given the strategy requiring to modify at one time both Vg1 and Vg2, a certain level of degeneracy (different bias pairs can determine different performance but same drain current) is evident in the two dimensional correlation plot in Fig.~\ref{fig19-c1}. 

\begin{figure}[h!]
         \resizebox{\hsize}{!}{\includegraphics{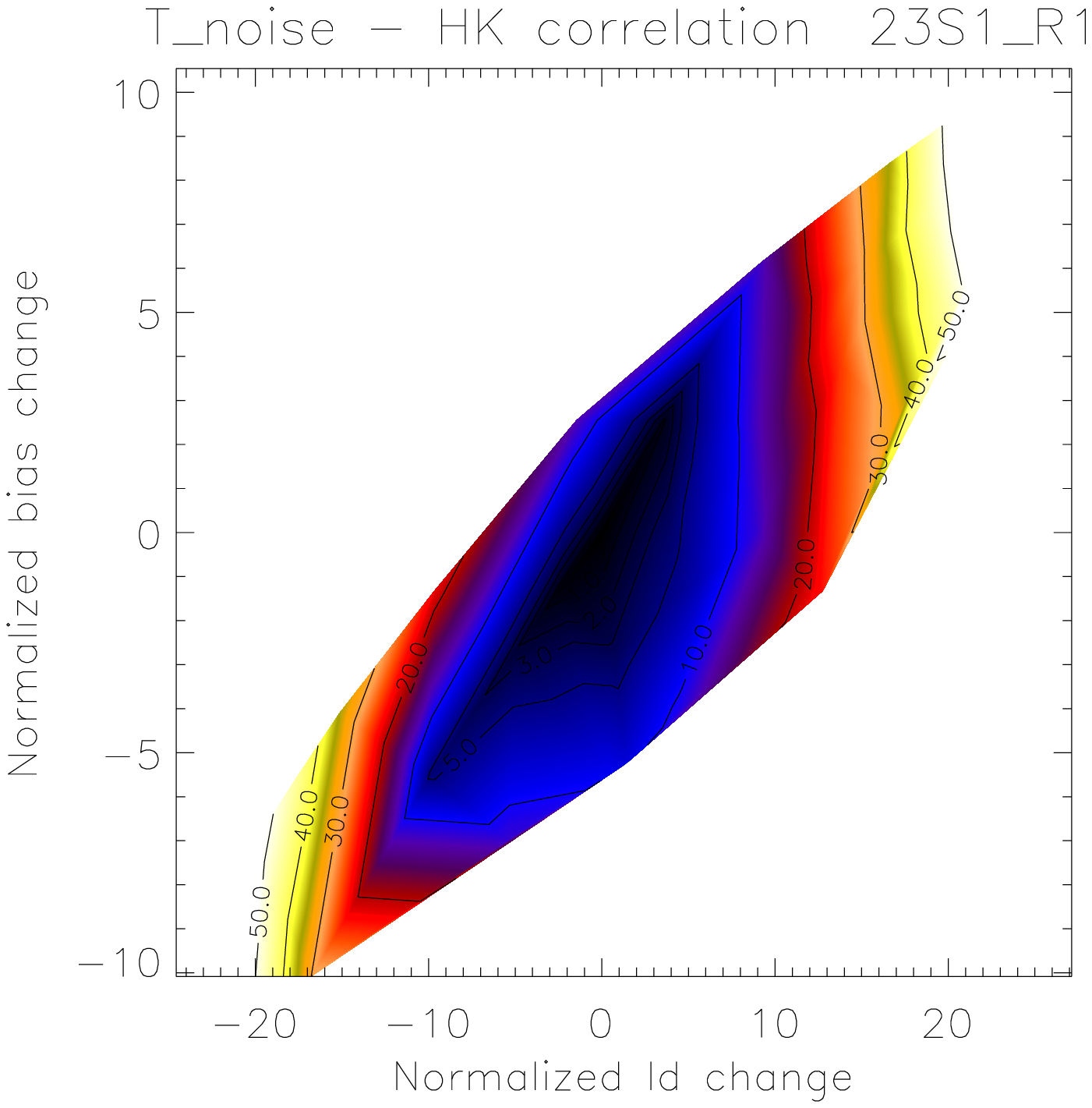}}
     		    \caption{Contour plot showing correlation between LNAs HK ( Id current and Bias Voltages) and Noise Temperature normalized to the best value: such a representation is able to break the degeneracy of the previous plot. On X axis the quantity $Id$/$Id_T$$\cdot$100, with $Id_T$ the drain current corresponding to the optimal tuned bias $Vg1_0$ , $Vg2_0$, and on Y axis the quantity [(Vg1 + Vg2)-($Vg1_0$ + $Vg2_0$)] / ($Vg1_0$ + $Vg2_0$) $\cdot$100, are represented. Dark regions refer to the best noise temperature points.
         		 }
         			\label{fig70-c}
     \end{figure}

Degeneration can be simply broken by properly adding the third dimension, giving also count of the total gate voltage deviation from nominal (Fig.~\ref{fig70-c}). 
		      
            The detailed comparison between optimal bias found at the different tuning levels is summarized in Appendix~\ref{app:EX-SUP} (see Tab.~\ref{tab:LNA_Tuning_settings})

        \paragraph{In flight tuning}
        \label{INPUTS-CPV}
        
%
%

            The complete Hyper Matrix Tuning (including also a limited number of combinations tested at three different drain voltages) will be performed in flight during the 4~K cooler cooldown which is expected to follow a profile similar to that shown in CSL. During the CSL test campaign, the four bias runs were performed in different thermal conditions, because of the 4K cooler cooldown, providing the reference loads with a reference stage cooling from 22K to 4.5K in about 11 days. Thermal slopes along the four steps ranged from 0.04K/h to 0.09K/h. These non steady thermal conditions increase the uncertainty in determining the absolute Noise Temperature and Isolation. However, the effect is small enough to provide an accurate estimation of the optimal bias corresponding to the minima).\\In order to optimally constrain the bias space (excluding configurations characterised by poor performance) a pre-tuning test has been devised to be run when the 4~K reference loads are at $\sim$25~K and the receivers observe a naturally imbalanced signal ($\sim 3$~K from the sky and $\sim 25$~K from the reference loads).
            
            This imbalance allows us to calculate an $Y$$^{*}$ factor (see Eq.~(\ref{eq:y_factor_pretuning})) slightly different with respect to Eq.~(\ref{eq:y_factor}), without actually changing the input temperature, and therefore provides a means to broadly identify regions where to concentrate the bias scans.

            \begin{eqnarray}
                &&\frac{T_{\rm ref}+T_n}{T_{\rm sky}+T_n}\approx \frac{V_{\rm ref}}{V_{\rm sky}}= Y^{*}\nonumber \\
                \mbox{ } \label{eq:y_factor_pretuning}\\
                && T_n \approx {\frac{T_{\rm ref}-Y^{*} \times T_{\rm sky}}{Y^{*}-1}}\nonumber
            \end{eqnarray}
        
\begin{figure*}
                \begin{center}
                    \includegraphics[width=7.5cm]{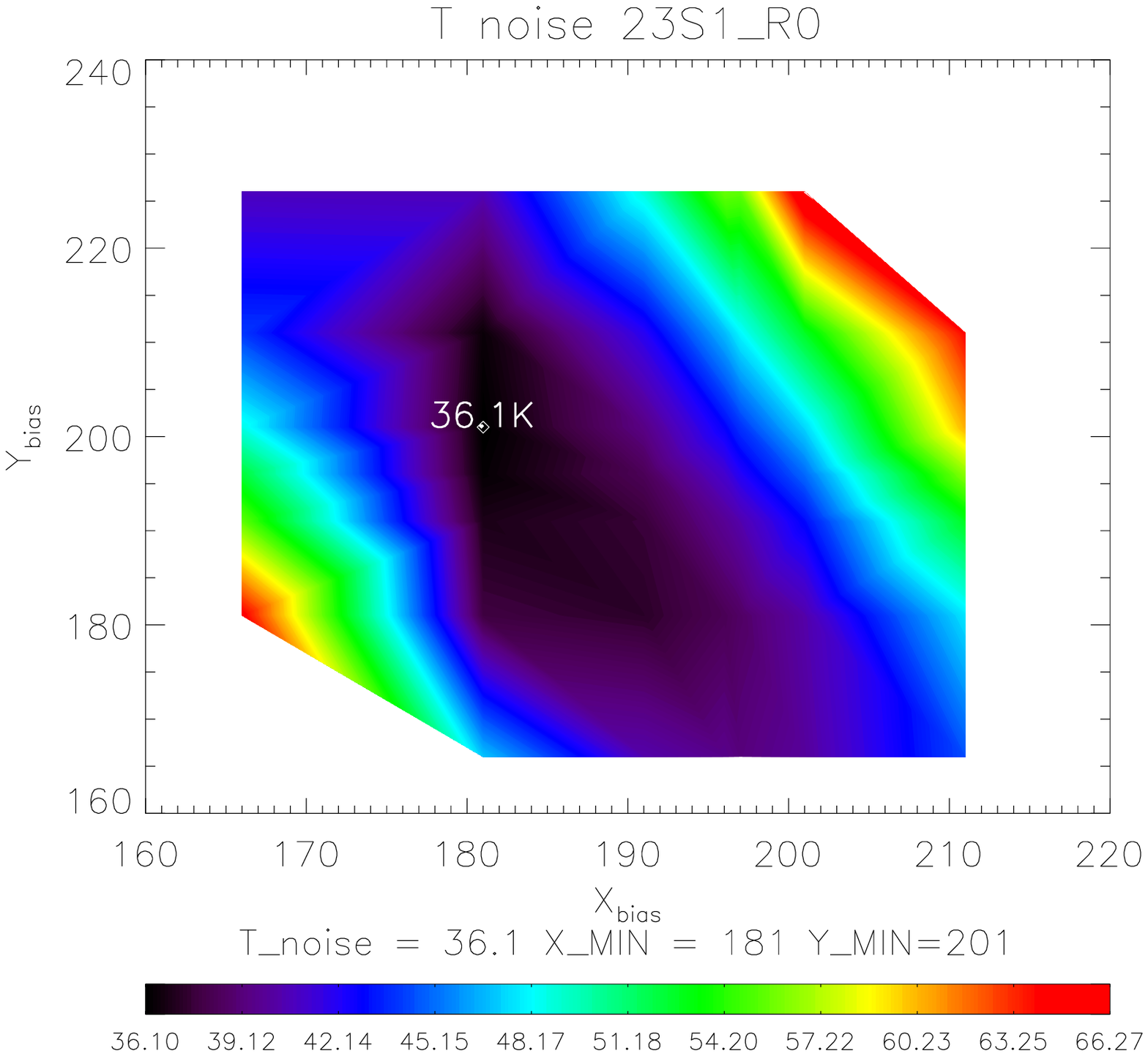}
                    \includegraphics[width=7.5cm]{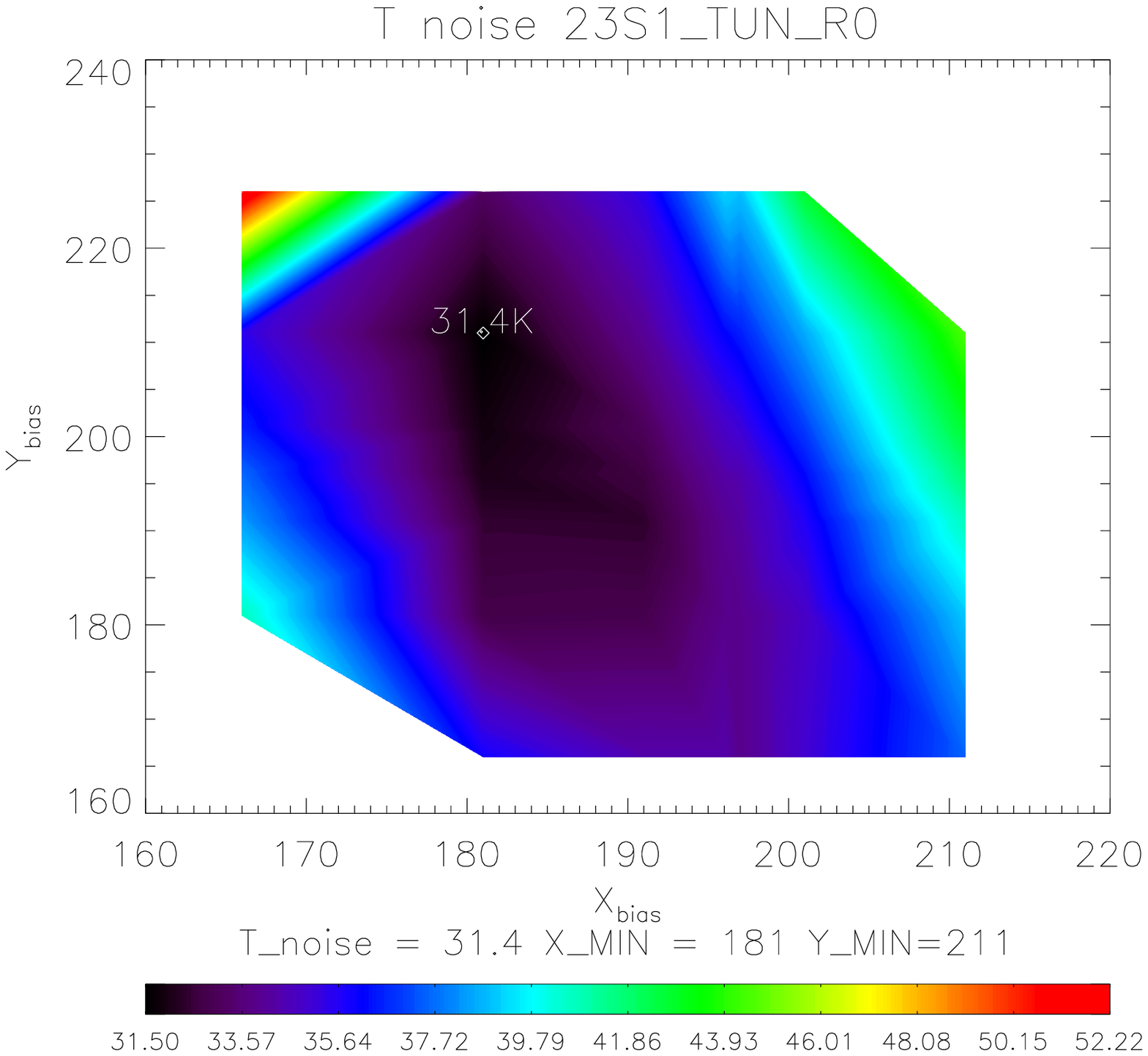}\\
                    \caption{
                        Matrix tuning results obtained with the ``Pretuning'' (left panel) and standard (right panel) tuning methods during satellite-level tests. The comparison shows that it is possible to identify at first order the best performance region with a single bias sweep, despite absolute noise temperatures can be different due to asymmetries' in the two legs; differences reduce when the thermal offset sky - ref increases, even if possible non linear effects can get stronger. In the case presented here, the offset was about 20K.
                    }
                    \label{fig23-c2}
                \end{center}            
            \end{figure*}

            In Fig.~\ref{fig23-c2} we show an example of tuning results obtained during satellite level tests using the intrinsic signal unbalance when the reference load was at $\sim 25$~K (left panel) and using the full dataset acquired during the cooldown (right panel).
            
            The main benefit of the Hyper Matrix Tuning strategy consists in its capability to optimizing at one time both Noise Temperature and Isolation; in fact, the measured performance correspond to the bias quadruplet that will be eventually chosen (while the pure 'matrix' scheme requires us to separately optimize noise temperature on each ACA and, only at the end, to mix bias pairs into quadruplets: such a scheme is intrinsically unable to measuring Isolation and also you do not know the phase of the individual tuning when you combine them).
		
		More benefits are related with the larger 4-dimensional bias space (many gate voltage combinations are included), and with the fact that possible electric drops due to coupling in paired ACAs, when they are tuned separately per gate voltage pairs, are naturally considered when changing bias in quadruplets and the amplifiers are naturally combined with the correct phases.
		
		Last but not least, with the Hyper Matrix strategy the level of signals collected by back end amplifiers and diodes is the same during the tuning as in nominal conditions after setting the optimal bias. With only one ACA on, the signal level is half the nominal; in the matrix scheme, bias are separately optimized in the two coupled ACAs, implying that the signal level changes when the two optimal bias pairs are combined to provide the optimal quadruplet. Both solutions could provide inaccurate results, due respectively to possible non linear response of the radiometers and to the bias cross talk between the paired ACAs.\\ 
		
		Full results are provided in Appendix~\ref{app:EX-SUP} (see Tab.~\ref{tab:LFI_TN_ISO_Tuning_comparison})

\begin{figure*}
                \begin{center}
                    \includegraphics[width=7.5cm]{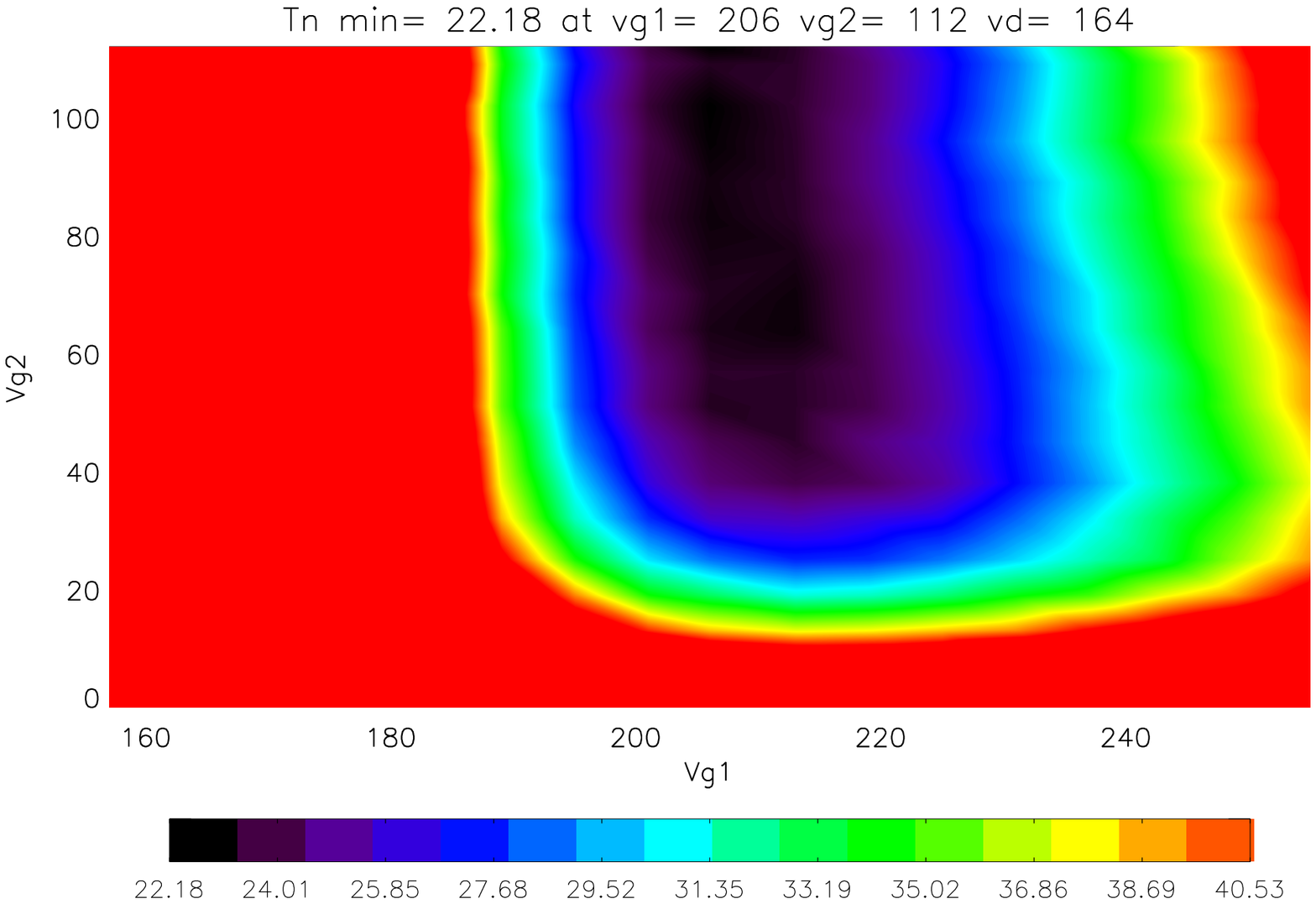}
                    \includegraphics[width=7.5cm]{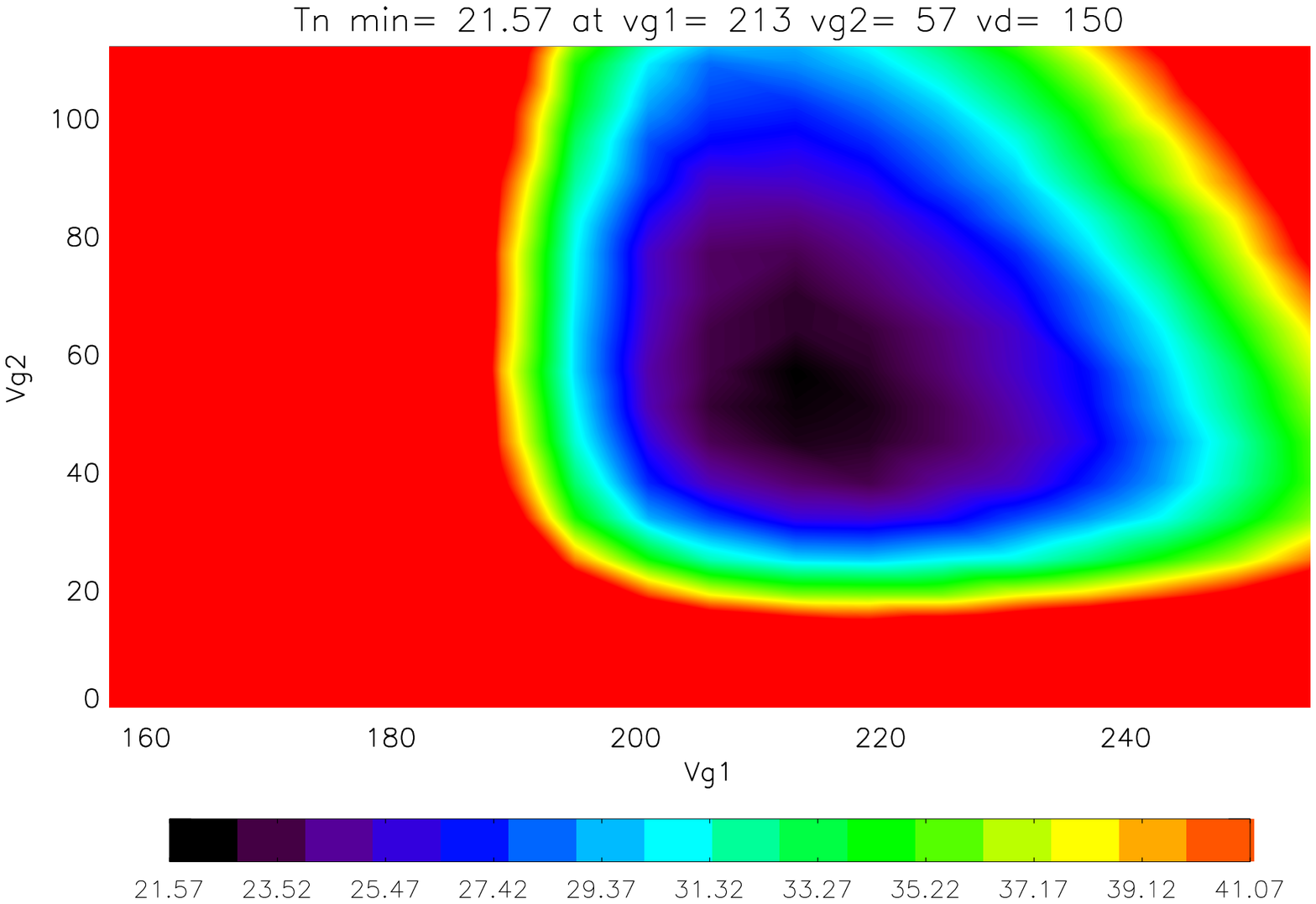}\\
                    \caption{
                        Vd change from nominal 164 DEC ( left panel) to 150 DEC (right panel) during FS 30 GHz matrix tuning. Noise Temperature seems to improve , although slightly (about 0.6K).
         Dedicated tests performed on Flight Spare Units (2008) confirmed that Vd Tuning can provide a further improvement.
                    }
                    \label{fig25-c}
                \end{center}            
            \end{figure*}

 \paragraph{Drain Voltage Tuning}
Dedicated tests were performed on Flight-Spare units (2008) to seek for further improvements in performance  by tuning also the drain voltage.\\
Results confirmed that Vd Tuning can improve noise (Fig.\ref{fig25-c}) and Isolation performance. During CPV Tuning, it is planned to be run only for three Vd values per ACA, changed with a matrix scheme on the two paired ACAs,  over a subset of fifteen gate voltage quadruplets individuated by applying the Pre-Tuning scheme. Hence, the bias space scanned becomes six-dimensional, at least for those few combinations expected to provide the best performance.\\

\section{Back-end electronics}
\label{DAE-TUN}

    \subsection{Analog signal processing}
    \label{DAE-THEO}

        The 44 analog voltage outputs from the radiometer back-end modules are digitised in the DAE box by 44 16-bit ADC converters. In particular the signal is processed as follows:
        \begin{equation}
            x\,\textrm{[V]} \rightarrow (x - V_0) \rightarrow (x - V_0) \times G_{\rm DAE}
            \rightarrow y\,\textrm{[14-bit value]},
            \label{eq:dae_offset_gain}
        \end{equation}
        where $V_0$ and $G_{\rm DAE}$ represent the offsets and gains which can be set independently for each channel. The voltage offset $V_0$ ranges from $\sim 0$~V to $\sim 2.5$~V and is programmable in 256 steps, while $G$ ranges from 1 to 48 in 10 steps. 

        The optimization is simple, as $V_0$ is chosen to make the signal average slightly above 0~Volt and $G_{DAE}$ is chosen to use about 75\%-80\% of the full ADC dynamic range. The best values can therefore be derived analytically from the output mean and standard deviation.
        
        The values of $V_0$ and $G$ are recorded into the housekeeping telemetry, so that the signal at the output of the back-end module can be reconstructed during ground analysis by inverting the relationship in Eq.~(\ref{eq:dae_offset_gain}).

        Optimal offset and gain parameters that resulted in during satellite-level tests (i.e. in the most flight-representative conditions) are listed in Table~\ref{tab:offset_gain_values}.

        \begin{table}[h!]
            \begin{center}
                \caption{Back-end voltage offset and gain optimal parameters. 44 GHz channels required the largest DAE Gain, because of their very low voltage output}
                \label{tab:offset_gain_values}
                \begin{tabular}{l c c c c}
                \hline
                \hline
                &\multicolumn{2}{c}{\textbf{M-00}}  
                &\multicolumn{2}{c}{\textbf{M-01}}  \\
                   &Gain   &Offset (V) &Gain   &Offset (V)\\
                        \hline
                LFI18   &1  &1.5  &1  &2.0    \\
                LFI19   &2  &1.0  &2  &1.0    \\
                LFI20   &2  &1.0  &2  &1.0  \\
                LFI21   &3  &0.5  &3  &0.5    \\
                LFI22   &6  &0.3  &6  &0.3    \\
                LFI23   &2  &0.8  &2  &0.8    \\
                LFI24   &24 &0.0  &24 &0.0  \\
                LFI25   &8  &0.0  &8  &0.0  \\
                LFI26   &16 &0.0  &12 &0.0  \\
                LFI27   &3  &0.5  &3  &0.5    \\
                LFI28   &3  &0.5  &3  &0.5    \\
                        \hline
                        \end{tabular}
                        
                        \begin{tabular}{l c c c c }
                        \hline
                &\multicolumn{2}{c}{\textbf{S-10}}  
                &\multicolumn{2}{c}{\textbf{S-11}}  \\
                        &Gain   &Offset (V) &Gain   &Offset (V)\\
                        \hline
                LFI18 &3  &0.6  &3  &0.6    \\
                LFI19 &4  &0.5  &4  &0.5    \\
                LFI20 &2  &1.0  &2  &1.0  \\
                LFI21 &2  &0.9  &2  &0.9    \\
                LFI22 &6  &0.3  &6  &0.3    \\
                LFI23 &2  &0.8  &4  &0.5    \\
                LFI24 &16 &0.0  &16 &0.0  \\
                LFI25 &8  &0.0  &12 &0.0  \\
                LFI26 &8  &0.0  &8  &0.0  \\
                LFI27 &3  &0.5  &3  &0.5    \\
                LFI28 &3  &0.5  &4  &0.5    \\
                \hline
                \end{tabular}
            \end{center}
        \end{table}

\paragraph{DAE NO-FLY ZONE} 

During the RAA tests we discovered an unexpected behaviour of the DAE. When exercising various values for $V_0$, it is expected that the error in the reconstructed signal $\epsilon = x - \tilde x$ exhibits small fluctuations, and that $\epsilon_\mathrm{sky} = \epsilon_\mathrm{ref}$, as sky and the reference signals are acquired at the same time and pass through the same detection chain. As a consequence, the difference $\left|\tilde x_\mathrm{sky} - \tilde x_\mathrm{ref}\right|$ should remain the same for different values of $V_0$, if the input loads do not change significantly.

	Instead, we discovered that for some well localized values of $V_0$ the value of $\left|\tilde x_\mathrm{sky} - \tilde x_\mathrm{ref}\right|$ shows sudden jumps. These effect varies with the voltage level of the ADC and disappears when sky and reference inputs are perfectly balanced. 

\begin{figure}
  \resizebox{\hsize}{!}{\includegraphics {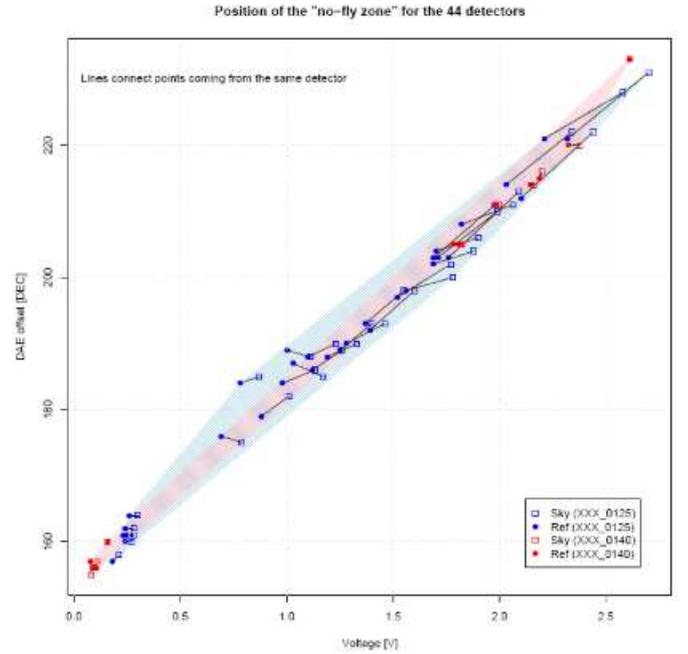}}
   \caption{The critical values for $V_0$ where the differential effect evidence is limited to a narrow region (nicknamed ``no fly zone'') when plotted against the absolute output voltage $\tilde x$ (X axis) and the value of $V_0$ in decimal units (Y axis). Red and blue points denote the result of cryogenic tests (performed at $\sim$ 20\,K in September 2006) and warm tests (performed at $\sim$ 300\,K in October 2006). Sky and reference samples are distinguished by their shape (empty squares for the sky, filled circles for the reference). Note that the 44 ADCs show a remarkable tendency to lie along the same line, and that the line is the same for warm and cryogenic tests.}
  \label{fig:fig27-c} 
 \end{figure}

 The root cause of this effect has not been understood. However we have carefully characterised the affected offset regions for each channel; these regions (called ``no fly zone'', i.e.\ the set of values of $V_0$, as function of the absolute signal unbalancing, shown in Fig.~\ref{fig:fig27-c}) that cover less than
10$\%$ of the whole offset range, will be avoided during instrument operation.  
 

\subsection{Digital signal processing}
\label {REBA-TUN}

        At the full sampling rate of 8192 samples/sec the output data rate from all scientific channels of the 14-bit ADC is of the order of 575 Mbps, largely exceding the telemetry requirement of 53.5 Kbps (including about ten Kbps reserved to housekeeping and diagnostic telemetry). Digital signal compression is therefore necessary to be able to download the produced telemetry to Earth \citep[see][]{2009_LFI_cal_M2}. This compression is achieved in three steps carried out by the Signal Processing Unit in the REBA box \citep[see][]{2009_LFI_cal_REBA}):

        \begin{itemize}
            \item data downsampling from 8192 Hz to $\sim$100~Hz (the actual sampling rate is programmable and dependent on the frequency channel);
            \item a lossy digital signal quantisation step;
            \item a lossless compression step.
        \end{itemize}

        The lossless compression performance, in particular, is critically dependent on the signal noise statistics, that need to be as close as possible to white noise. This is achieved by transforming the sky and reference load data streams (which are characterised by a strong 1/$f$ noise component) into two differential data streams according to the following formula:

        \begin{equation}
            (V_{\rm sky}, V_{\rm ref})\rightarrow(V_1, V_2),
            \label{eq:reba_differencing}
        \end{equation}
        where
        
        \begin{eqnarray}
            V_1 &=& V_{\rm sky} - r_1\, V_{\rm ref} \nonumber\\
            \mbox{ }\\
            V_2 &=& V_{\rm sky} - r_2\, V_{\rm ref}, \nonumber
            \label{eq:reba_differencing_2}
        \end{eqnarray}
        which greatly reduces the 1/$f$ fluctuations. The parameters $r_1$ and $r_2$, that can be controlled and uploaded into memory, must be tuned in order to ensure that the on-ground reconstruction of the total power data streams is not affected by errors.

        The two differential data streams are then quantised according to the following formula:

        \begin{equation}
            Q_i= {\rm round}\left[(V_i + \Offset)\times \SecondQuant\right], i=1,2,
            \label{eq:quantisation}
        \end{equation}
        where $\Offset$ and $\SecondQuant$ are an offset and a quantisation factor. Finally the down-sampled and quantised differential data streams are then compressed by a lossless compression algorithm before building telemetry packets. 

        The optimal set of parameters \citep[see][]{2009_LFI_cal_D2} $r_1$, $r_2$, $\Offset$ and $\SecondQuant$ can be found in two phases: (i) first the data are acquired unquantized and uncompressed and the optimal parameters are found by applying a software model of the Signal Processing Unit to these data with different sets of quantisation parameters; (ii) then the optimal parameters are uploaded to the instrument and data are acquired in quantised and compressed mode in order to verify that the desired compression rate is met and that the scientific quality loss is negligible.

        In Table~\ref{tab:reba_values} we list the optimal parameters found during satellite-level tests. With those values the telemetry constraints have been largely met with a minimal loss in signal quality, obtaining a ratio between the noise standard deviation $\sigma$ and the quantisation step, $q$ well within the scientific requirement $\sigma / q \gtrsim 2$.
        
 
        \begin{table}[h!]
        \begin{center}
            \caption{Digital quantisation optimal parameters.}
            \label{tab:reba_values}
            \begin{tabular}{l c c c c c c}
                \hline
                \hline
                \textbf{Channel}& $r_1$ & $r_2$ & $s_q$ & $C_r$ & $\Qerrdiff$ \\
                \hline

1800	&	1.042	&0.958	&4.073 	&	2.390	&	0.0358	\\
1801	&	1.042	&0.958	&3.161 	&	2.380	&	0.0348	\\
1810	&	1.042	&0.917	&2.765 	&	2.380	&	0.0352	\\
1811	&	1.042	&0.958	&3.651 	&	2.380	&	0.0357	\\
1900	&	1.042	&0.958	&3.727 	&	2.390	&	0.0356	\\
1901	&	1.042	&0.958	&3.374 	&	2.380	&	0.0343	\\
1910	&	1.042	&0.958	&3.265 	&	2.370	&	0.0346	\\
1911	&	1.042	&0.958	&2.767 	&	2.350	&	0.0337	\\
2000	&	1.042	&0.958	&3.413 	&	2.380	&	0.0343	\\
2001	&	1.042	&0.958	&3.130 	&	2.360	&	0.0342	\\
2010	&	1.042	&0.958	&3.446 	&	2.380	&	0.0346	\\
2011	&	1.042	&0.958	&3.529 	&	2.380	&	0.0347	\\
2100	&	1.042	&0.958	&4.120 	&	2.400	&	0.0350	\\
2101	&	1.042	&0.958	&4.377 	&	2.400	&	0.0362	\\
2110	&	1.042	&0.958	&4.293 	&	2.410	&	0.0360	\\
2111	&	1.042	&0.958	&4.050 	&	2.400	&	0.0354	\\
2200	&	1.042	&1.000	&3.684 	&	2.340	&	0.0341	\\
2201	&	1.042	&1.000	&3.345 	&	2.330	&	0.0320	\\
2210	&	1.083	&1.000	&3.085 	&	2.360	&	0.0352	\\
2211	&	1.083	&1.000	&2.603 	&	2.340	&	0.0339	\\
2300	&	1.042	&1.000	&5.096 	&	2.380	&	0.0359	\\
2301	&	1.042	&0.958	&4.366 	&	2.410	&	0.0363	\\
2310	&	1.042	&0.958	&4.585 	&	2.410	&	0.0360	\\
2311	&	1.042	&0.958	&4.098 	&	2.400	&	0.0355	\\
  \hline
2400	&	1.042	&0.917	&4.375 	&	2.430	&	0.0372	\\
2401	&	1.083	&0.875	&3.374 	&	2.430	&	0.0362	\\
2410	&	1.042	&0.917	&4.490 	&	2.430	&	0.0356	\\
2411	&	1.042	&0.917	&4.827 	&	2.440	&	0.0366	\\
2500	&	1.042	&0.917	&6.351 	&	2.450	&	0.0382	\\
2501	&	1.042	&0.917	&6.699 	&	2.460	&	0.0380	\\
2510	&	1.000	&0.958	&5.698 	&	2.390	&	0.0349	\\
2511	&	1.042	&0.917	&5.289 	&	2.440	&	0.0372	\\
2600	&	1.000	&0.917	&4.571 	&	2.420	&	0.0357	\\
2601	&	1.042	&0.875	&4.824 	&	2.440	&	0.0372	\\
2610	&	1.042	&0.917	&5.866 	&	2.450	&	0.0369	\\
2611	&	1.000	&0.958	&7.496 	&	2.430	&	0.0349	\\
  \hline
2700	&	1.042	&0.833	&3.099 	&	2.410	&	0.0328	\\
2701	&	1.000	&0.085	&2.930 	&	2.390	&	0.0307	\\
2710	&	1.000	&0.875	&3.299 	&	2.400	&	0.0351	\\
2711	&	1.042	&0.875	&3.438 	&	2.420	&	0.0362	\\
2800	&	1.083	&1.000	&3.634 	&	2.380	&	0.0338	\\
2801	&	1.083	&1.000	&2.984 	&	2.350	&	0.0327	\\
2810	&	1.042	&0.958	&4.130 	&	2.400	&	0.0349	\\
2811	&	1.042	&0.958	&3.583 	&	2.380	&	0.0341	\\

                     \hline
                    \hline
            \end{tabular}
        \end{center}
    \end{table}

\section{Conclusions}
\label {CONCLUSIONS}
	
    The Planck-LFI instrument scientific performance critically depends on a number of parameters that need to be tuned before starting nominal operations.
    
    Bias voltages and currents to front end amplifiers represent the most critical parameters as they determine the final receiver sensitivity and isolation. Due to the complex instrument grounding and thermal distribution these biases need to be tuned every time environmental conditions change, and therefore a tuning activity has been performed at each stage of the test campaign. In this paper we have shown how the bias tuning strategy has evolved in time up to the current strategy that is about to be applied during flight calibration and foresees a complex scheme to scan efficiently the 4-dimensional bias parameter space.
    
    Also the back-end analog and digital electronic units need to be tuned in order to optimise the signal scientific quality. In particular the receiver output voltage must be adapted to the ADC dynamic range using a programmable gain/offset stage and after digitisation the signal must be further quantised and compressed to comply with the available telemetry bandwidth. We have shown how all these parameters can be optimised through dedicated tests and we have presented the most up to date parameters obtained during satellite-level tests.
    
    At the time of writing the satellite is approaching L2 and the in-flight calibration phase has just begun. This will be a critical time in which the final tuning will be performed and the instrument scientific performance will be definitely set.

\vspace{4mm}
\begin{acknowledgements}
"Planck is a project of the European Space Agency with instruments
funded by ESA member states, and with special contributions from Denmark
and NASA (USA). The Planck-LFI project is developed by an International
Consortium lead by Italy and involving Canada, Finland, Germany, Norway,
Spain, Switzerland, UK, USA. The Italian contribution to Planck is
supported by the Italian Space Agency (ASI)."
\end{acknowledgements}

\bibliographystyle{aa} 
\bibliography{Cuttaia}

\hyphenation{Post-Script Sprin-ger}
\begin{thebibliography}{27}
\expandafter\ifx\csname natexlab\endcsname\relax\def\natexlab#1{#1}\fi

\bibitem[{{Artal} {et~al.}(2009){Artal}, {Aja}, {L. de la Fuente}, {Pascual},
  {Mediavilla}, {Martinez-Gonzalez}, {Pradell}, {de Paco}, {Bara }, {Blanco},
  {Garc{\'i}a }, {Davis }, {Kettle}, {Roddis}, {Wilkinson }, {Bersanelli},
  {Mennella }, {Tomasi}, {Butler}, {Cuttaia}, {Mandolesi}, \&
  {Stringhetti}}]{2009_LFI_cal_R9}
{Artal}, E., {Aja}, B., {L. de la Fuente}, M., {et~al.} 2009, {JINST, this
  issue}, Submitted

\bibitem[{{Bersanelli} {et~al.}(2009){Bersanelli}, {Cappellini}, {Cavaliere},
  {Donzelli}, {Maino}, {Mennella}, {Pezzati}, {Tomasi}, {Zonca},
  {D’Arcangelo}, {Figini}, {Garavaglia}, {Platania}, {Simonetto}, {Sozzi},
  {Blackhurst}, {Davies}, {Davis}, {Edgeley}, {Galtress}, {Kettle}, {Lawson},
  {Leahy}, {Lowe}, {Roddis}, {Wilkinson}, {Winder}, {Bhandari}, {Bowman},
  {Gaier}, {Gorski}, {Janssen}, {Lawrence}, {Levin}, {Nash}, {Paine},
  {Partridge}, {Pearson}, {Prina}, {Seiffert}, {Wade}, {Smoot}, {Leonardi},
  {Lubin}, {Meinhold}, {Artina}, {Balasini}, {Baldan}, {Battaglia}, {Bastia},
  {Boschini}, {Cafagna}, {Colombo}, {Ferrari}, {Franceschet}, {Lapolla},
  {Leutenegger}, {Miccolis}, {Pagan}, {Pecora}, {Silvestri}, {Aja}, {Artal},
  {De La Fuente}, {Herranz}, {Mediavilla}, {Pascual}, {Bernardino},
  {Martinez-Gonzalez}, {Salmon}, {Vielva}, {Tauber}, {Bennett}, {Crone},
  {Marti-Canales}, {Peres-Cuevas}, {Mendes}, {Herreros}, {Hildebrandt},
  {Gomez}, {Hoyland}, {Rebolo}, {Rubino}, {Poutanen}, {Tuovinen}, {Varis},
  {Hughes}, {Jukkala}, {Kilpela}, {Laaninen}, {Sjoman}, {Burigana}, {Butler},
  {Cuttaia}, {De Rosa}, {Finelli}, {Franceschi}, {Malaspina}, {Mandolesi},
  {Morgante}, {Popa}, {Sandri}, {Stringhetti}, {Terenzi}, {Valenziano},
  {Villa}, {Nesti}, {Meharga}, {Morrisset}, {Rohlfs}, {T{\"u}rler}, {Binko},
  {Gregorio}, {Fogliani}, {Frailis}, {Galeotta}, {Gasparo}, {Maris}, {Pasian},
  {Zacchei}, {Perrotta}, {Manzato}, {Maggio}, {Natoli}, {De Gasperis},
  {Vittorio}, {Balbi}, {Baccigalupi}, {Danese}, {Stivoli}, {Leach}, {Lilje},
  {Banday}, {De Zotti}, {Courvoisier}, {Lahteenmaki}, {Matarrese},
  {Norgaard-Nielsen }, {Scott}, {Silk}, {Enqvist}, {Tofani}, {White}, {De
  Angelis}, \& {Falvella}}]{2009_LFI_cal_M2}
{Bersanelli}, M., {Cappellini}, B., {Cavaliere}, F., {et~al.} 2009, {\aap},
  Submitted

\bibitem[{Bhandari {et~al.}(2004)Bhandari, Prina, Bowman, Paine, Pearson, \&
  Nash}]{2004_Sorcio}
Bhandari, P., Prina, M., Bowman, R.~C., {et~al.} 2004, Cryogenics, 44, 395 ,
  2003 Space Cryogenics Workshop

\bibitem[{{Bradshaw} \& {Orlowska}(1997)}]{4K_cooler}
{Bradshaw}, T. \& {Orlowska}, A. 1997, {Proc. 6th European Symposium on Space
  Environmental Control Systems,ESA SP400}, V.2, 465

\bibitem[{{Cuttaia}(2005)}]{2005_Phd_Cuttaia}
{Cuttaia}, F. 2005, {Ph.D Thesis in Astronomy, Alma Mater Studiorum, Bologna}

\bibitem[{{Cuttaia} {et~al.}(2009){Cuttaia}, {Terenzi}, {Valenziano}, {Sandri},
  {Bersanelli}, {Battaglia}, {Colombo}, {Lapolla}, {Butler}, {De Rosa},
  {Mandolesi}, {Morgante}, {Stringhetti}, {Villa}, {Biggi}, {Lapini}, \&
  {Panagin}}]{2009_LFI_cal_T5}
{Cuttaia}, F., {Terenzi}, L., {Valenziano}, L., {et~al.} 2009, {JINST, this
  issue}, In preparation

\bibitem[{{Davis} {et~al.}(2009){Davis}, {Wilkinson}, {Davies}, {Winder},
  {Roddis}, {Blackhurst}, {Lawson}, {Lowe }, {Baines}, {Butlin}, {Galtress },
  {Shepherd }, {Aja}, {Artal}, {Bersanelli}, {Butler}, {Castelli }, {Cuttaia},
  {D’Arcangelo}, {Hoyland}, {Kettle}, {Leonardi }, {Mandolesi}, {Mennella },
  {Meinhold }, {Pospieszalski}, {Stringhetti}, {Tomasi}, {Valenziano}, \&
  {Zonca}}]{2009_LFI_cal_R8}
{Davis}, R., {Wilkinson}, A., {Davies}, R., {et~al.} 2009, {JINST, this issue},
  Submitted

\bibitem[{{Hoyland}(2003)}]{2003_hoyland_espoo}
{Hoyland}, R. 2003, in ESA Workshop on Millimetre Wave Technology and ich
  Applications, Vol. 211, Proceedings of 3rd ESA Workshop on Millimetre se Wave
  Technology and Applications, 305--310

\bibitem[{{Lamarre}(2009)}]{2009_HFI_PAPER}
{Lamarre}, J. 2009, {\aap}, Submitted to Astr.Astro.

\bibitem[{{Malaspina} \& {Franceschi}(2009)}]{2009_LFI_cal_DE}
{Malaspina}, M. \& {Franceschi}, E. 2009, {JINST, this issue}, Submitted

\bibitem[{{Mandolesi} {et~al.}(2009){Mandolesi}, {Aja}, {Artal}, {Artina},
  {Baccigalupi}, {Balasini}, {Balbi}, {Baldan}, {Banday}, {Bastia},
  {Battaglia}, {Bennett}, {Bernardino}, {Bersanelli}, {Bhandari}, {Binko},
  {Blackhurst}, {Boschini}, {Bowman}, {Bremer}, {Burigana}, {Butler},
  {Cafagna}, {Cappellini}, {Cavaliere}, {Colombo}, {Courvoisier}, {Crone},
  {Cuttaia}, {D'Arcangelo}, {Danese}, {Davies}, {Davis}, {De Angelis}, {De
  Gasperis}, {De La Fuente}, {De Rosa}, {De Zotti}, {Donzelli}, {Edgeley},
  {Enqvist}, {Ensslin}, {Falvella}, {Ferrari}, {Figini}, {Finelli}, {Fogliani},
  {Frailis}, {Franceschet}, {Franceschi}, {Gaier}, {Galeotta}, {Galtress},
  {Garavaglia}, {Gasparo}, {Giardino}, {Gomez}, {Gorski}, {Gregorio}, {Hazell},
  {Hell}, {Herranz}, {Herreros}, {Hildebrandt}, {Hovest}, {Hoyland}, {Hughes},
  {Janssen}, {Jukkala}, {Kettle}, {Kilpel{\"a}}, {Laaninen}, {Lahteenmaki},
  {Lapolla}, {Lawrence}, {Lawson}, {Leach}, {Leahy}, {Leonardi}, {Leutenegger},
  {Levin}, {Lilje}, {Lowe}, {Lubin}, {Maggio}, {Maino}, {Malaspina}, {Manzato},
  {Maris}, {Marti-Canales}, {Martinez-Gonzalez}, {Matarrese}, {Matthai},
  {Mediavilla}, {Meharga}, {Meinhold}, {Mendes}, {Mennella}, {Miccolis},
  {Morgante}, {Morrisset}, {Nash}, {Natoli}, {Nesti}, {Norgaard-Nielsen},
  {Pagan}, {Paine}, {Partridge}, {Pascual}, {Pasian}, {Pearson}, {Pecora},
  {Peres-Cuevas}, {Perrotta}, {Pezzati}, {Phipps}, {Platania}, {Popa},
  {Poutanen}, {Prina}, {Rachen}, {Rebolo}, {Reinecke}, {Riller}, {Roddis},
  {Rohlfs}, {Rubino}, {Salmon}, {Sandri}, {Scott}, {Seiffert}, {Silk},
  {Silvestri}, {Simonetto}, {Sjoman}, {Smoot}, {Sozzi}, {Sternberg}, {Stivoli},
  {Stringhetti}, {Tauber}, {Terenzi}, {Tofani}, {Tomasi}, {Tuovinen},
  {T{\"u}rler}, {Uwe}, {Valenziano}, {Varis}, {Vielva}, {Villa}, {Vittorio},
  {Wade}, {White}, {Wilkinson}, {Winder}, {Zacchei}, \&
  {Zonca}}]{2009_LFI_cal_M1}
{Mandolesi}, N., {Aja}, B., {Artal}, E., {et~al.} 2009, {\aap}, Submitted

\bibitem[{{Maris} {et~al.}(2009){Maris}, {Bersanelli}, {D'Arcangelo}, {Maino},
  {Mennella}, {Tomasi}, {Zonca}, {Lowe}, {Leonardi}, {Meinhold}, {Miccolis},
  {J. Salmon}, {Mendes}, {Herreros}, {Hildebrandt}, {Butler}, {Burigana},
  {Cuttaia}, {Franceschi}, {Malaspina}, {Mandolesi}, {Morgante}, {Sandri},
  {Terenzi}, {Valenziano}, {Villa}, {Binko}, {Meharga}, {Morrisset}, {Rohlfs},
  {Turler}, {Fogliani}, {Frailis}, {Galeotta}, {Gasparo}, {Gregorio}, {Maggio},
  {Manzato}, {Pasian}, {Perrotta}, \& {Zacchei}}]{2009_LFI_cal_D2}
{Maris}, M., {Bersanelli}, M., {D'Arcangelo}, O., {et~al.} 2009, {JINST, this
  issue}, Submitted

\bibitem[{{Meinhold} {et~al.}(2009){Meinhold}, {Leonardi}, {Artal},
  {Battaglia}, {Bersanelli}, {Blackhurstis}, {Butler}, {Cuevas}, {Cuttaia},
  {D’Arcangelo}, {Davis}, {Frailis}, {Franceschet}, {Franceschi}, {Gaier},
  {Galeotta}, {Gregorio}, {Hoyland}, {Hughes}, {Jukkala}, {Kettle}, {Laaninen},
  {Leahy}, {Leutenegger}, {Lowe}, {Malaspina}, {Mandolesi}, {Maris},
  {Martinez}, {Mendes}, {Mennella}, {Miccolis}, {Morgante}, {Roddis}, {Sandri},
  {Seiffert}, {Salmon}, {Stringhetti}, {Poutanen}, {Terenzi}, {Tomasi},
  {Tuovinen}, {Varis}, {Valenziano}, {Villa}, {Wilkinson}, {Winder}, {Zacchei},
  \& {Zonca}}]{2009_LFI_cal_R2}
{Meinhold}, P., {Leonardi}, R., {Artal}, E., {et~al.} 2009, {JINST, this issue}

\bibitem[{{Mennella} {et~al.}(2009){Mennella}, {Bersanelli}, {Aja}, {Artal},
  {Artina}, {Balasini}, {Baldan}, Bastia, {Battaglia}, {Bernardino},
  {Boschini}, {Butler}, {Cafagna}, {Cappellini}, {Cavaliere}, {Colombo},
  {Cuttaia}, {D'Arcangelo}, {Davis}, {de La Fuente}, {Edgeley}, {Falvella},
  {Ferrari}, {Fogliani}, {Frailis}, {Franceschet}, {Franceschi}, {Gaier},
  {Galeotta}, {Gomez}, {Gregorio}, {Herreros}, {Hildebrandt}, {Hoyland},
  {Hughes}, {Jukkala}, {Kettle}, {Laaninen}, {Lapolla}, {Lawrence}, {Lawson},
  {Leach}, {Lehay}, {Leonardi}, {Leutenegger}, {Levin}, {Lilje}, {Lowe},
  {Maino}, {Malaspina}, {Mandolesi}, {Manzato}, {Maris}, {Marti-Canales},
  {Martinez-Gonzalez}, {Mediavilla}, {Meinhold}, {Mendes}, {Miccolis},
  {Morgante}, {Nesti}, {Pagan}, {Pasian}, {Pasqual}, {Pecora}, {Peres-Cuevas},
  {Platania}, {Pospieszalsky}, {Poutanen}, {Rebolo}, {Roddis}, {Rubino},
  {Salmon}, {Sandri}, {Seiffert}, {Silvestri}, {Simonetto}, {Sozzi},
  {Stringhetti}, {Tauber}, {Terenzi}, {Tomasi}, {Tuovinen}, {Valenziano},
  {Varis}, {Villa}, {Wilkinson}, {Winder}, {Zacchei}, \&
  {Zonca}}]{2009_LFI_cal_M3}
{Mennella}, A., {Bersanelli}, M., {Aja}, B., {et~al.} 2009, {\aap}, Submitted

\bibitem[{{Mennella} {et~al.}(2003){Mennella}, {Bersanelli}, {Seiffert},
  {Kettle}, {Roddis}, {Wilkinson}, \& {Meinhold}}]{mennella03}
{Mennella}, A., {Bersanelli}, M., {Seiffert}, M., {et~al.} 2003, {\aap}, 410,
  1089

\bibitem[{{Morgante} {et~al.}(2009){Morgante}, {Pearson}, \&
  {Stassi}}]{2009_LFI_cal_T2}
{Morgante}, G., {Pearson}, D., \& {Stassi}, P., e.~a. 2009, {JINST, this
  issue}, Submitted

\bibitem[{{Rebolo} \& {Herreros}(2009)}]{2009_LFI_cal_REBA}
{Rebolo}, R. \& {Herreros}, J. 2009, {JINST, this issue}, Submitted

\bibitem[{{Seiffert} {et~al.}(2002){Seiffert}, {Mennella}, {Burigana},
  {Mandolesi}, {Bersanelli}, {Meinhold}, \& {Lubin}}]{Seiffert02}
{Seiffert}, M., {Mennella}, A., {Burigana}, C., {et~al.} 2002, {\aap}, 391,
  1185

\bibitem[{{Tauber}(2009)}]{2009_Tauber_Planck_Optics}
{Tauber}, J. 2009, {\aap}, Submitted

\bibitem[{{Terenzi} {et~al.}(2009{\natexlab{a}}){Terenzi}, {Bersanelli},
  {Mennella}, {Tomasi}, {Battaglia}, {Lapolla}, {Hughes}, {Kilpelä},
  {Laaninen}, {Butler}, {De Rosa}, {Franceschi}, {Mandolesi}, {Morgante},
  {Stringhetti}, {Valenziano}, {Galeotta}, {Maris}, \&
  {Zacchei}}]{2009_LFI_cal_T1}
{Terenzi}, L., {Bersanelli}, M., {Mennella}, A., {et~al.} 2009{\natexlab{a}},
  {JINST, this issue}, Submitted

\bibitem[{{Terenzi} {et~al.}(2009{\natexlab{b}}){Terenzi}, {Cuttaia}, {De
  Rosa}, {Lapolla}, {Sandri}, {Pecora}, {Biggi}, {Lapini}, {Panagin},
  {Bersanelli}, {Battaglia}, {Butler}, {Mandolesi}, {Morgante}, {Valenziano},
  \& {Zacchei}}]{2009_LFI_cal_T4}
{Terenzi}, L., {Cuttaia}, F., {De Rosa}, A., {et~al.} 2009{\natexlab{b}},
  {JINST, this issue}, In preparation

\bibitem[{{Tomasi} {et~al.}(2009){Tomasi}, {Mennella}, {Galeotta}, {Lowe},
  {Mendes}, {Leonardi}, {Villa}, {Cappellini}, {Gregorio}, {Meinhold},
  {Sandri}, {Cuttaia}, {Terenzi}, {Maris}, {Valenziano}, Salmon, {Bersanelli},
  {Binko}, {Butler}, {D'Arcangelo}, {Fogliani}, {Frailis}, {Franceschi},
  {Gasparo}, {Maggio}, {Maino}, {Malaspina}, {Mandolesi}, {Manzato}, {Meharga},
  {Morgante}, {Morrisset}, {Pasian}, {Perrotta}, {Rohlfs}, {t\"urler},
  {Zacchei}, \& {Zonca}}]{2009_LFI_cal_D4}
{Tomasi}, M., {Mennella}, A., {Galeotta}, S., {et~al.} 2009, {JINST, this
  issue}, Submitted

\bibitem[{{Valenziano} {et~al.}(2009){Valenziano}, {Cuttaia}, {De Rosa},
  {Terenzi}, {Brighenti}, {Cazzola}, {Garbesi}, {Mariotti}, {Pagan}, {Orsi},
  {Battaglia}, {Butler}, {Bersanelli}, {D'Arcangelo}, {Levin}, {Mandolesi},
  {Mennella}, {Morgante}, {Morigi}, {Sandri}, {Simonetto}, {Tomasi}, {Villa},
  {Frailis}, {Galeotta}, {Gregorio,}, {Leonardi}, {Lowe}, {Maris}, {Meinhold},
  {Mendes}, {Stringhetti}, {Zonca}, \& {Zacchei}}]{2009_LFI_cal_R1}
{Valenziano}, L., {Cuttaia}, F., {De Rosa}, A., {et~al.} 2009, {JINST, this
  issue}

\bibitem[{{Varis} {et~al.}(2009){Varis}, {Hughes}, {Laaninen}, {Kilpi\"{a}},
  {Jukkala}, {Tuovinen}, {Ovaska}, {Sj\"{o}man}, {Kangaslahti}, {Gaier},
  {Hoyland}, {Meinhold}, {Mennella}, {Bersanelli}, {Butler}, {Cuttaia},
  {Leonardi}, {Leutenegger}, {Mandolesi}, {Miccolis}, {Poutanen},
  {Kurki-Suonio}, {Stringhetti}, {Tomasi}, \& {Valenziano}}]{2009_LFI_cal_R10}
{Varis}, J., {Hughes}, N., {Laaninen}, M., {et~al.} 2009, {JINST, this issue},
  Submitted

\bibitem[{{Villa} {et~al.}(2009{\natexlab{a}}){Villa}, {Bersanelli},
  {Cappellini}, {Cavaliere}, {Maino}, {Mennella }, {Tomasi}, {Zonca},
  {D'Arcangelo}, {Simonetto}, {Sozzi}, {Blackhurst}, {Davis}, {Edgeley},
  {Galtress }, {Kettle}, {Lawson}, {Leahy}, {Lowe }, {Roddis}, {Wilkinson},
  {Winder}, {Pospieszalski}, {Gaier}, {Lawrence }, {Leonardi }, {Meinhold },
  {Balasini }, {Battaglia}, {Boschini }, {Colombo}, {Franceschet}, {Lapolla},
  {Leutenegger}, {Marseguerra}, {Miccolis }, {Pagan}, {Pecora}, {Silvestri},
  {Aja}, {Artal}, {Colin}, {De La Fuente }, {Mediavilla}, {Pascual},
  {Bernardino}, {Martinez-Gonzalez}, {Salmon}, {Mendes}, {Hoyland}, {Poutanen
  }, {Tuovinen }, {Varis}, {Hughes}, {Jukkala}, {Kilpel{\"a}}, {Laaninen },
  {Sjoman}, {Butler}, {Cuttaia}, {Franceschi}, {Malaspina}, {Mandolesi},
  {Morgante }, {Sandri}, {Stringhetti}, {Terenzi}, {Valenziano}, {Gregorio },
  {Frailis}, {Galeotta }, {Maris}, {Pasian}, \& {Zacchei}}]{2009_LFI_cal_M4}
{Villa}, F., {Bersanelli}, M., {Cappellini}, B., {et~al.} 2009{\natexlab{a}},
  {\aap}, Submitted

\bibitem[{{Villa} {et~al.}(2009{\natexlab{b}}){Villa}, {D'Arcangelo}, {Pagana},
  {Figini}, {Garavaglia}, {Pecora}, {Simonetto}, {Sozzi}, {Battaglia},
  {Bersanelli}, {Butler}, \& {Mandolesi}}]{2009_LFI_cal_O2}
{Villa}, F., {D'Arcangelo}, O., {Pagana}, E., {et~al.} 2009{\natexlab{b}},
  {JINST, this issue}, Submitted

\bibitem[{{Villa} {et~al.}(2009{\natexlab{c}}){Villa}, {D'Arcangelo}, {Pecora},
  {Figini}, {Nesti}, {Simonetto}, {Sozzi}, {Sandri}, {Battaglia}, {Bersanelli},
  {Butler}, \& {Mandolesi}}]{2009_LFI_cal_O1}
{Villa}, F., {D'Arcangelo}, O., {Pecora}, M., {et~al.} 2009{\natexlab{c}},
  {JINST, this issue}, Submitted

\end{thebibliography}

\appendix

\section{LFI receiver and channel naming convention}
\label{app:naming_convention}

\begin{figure}[h!]
   \begin{center}
      \includegraphics[width=7.5cm]{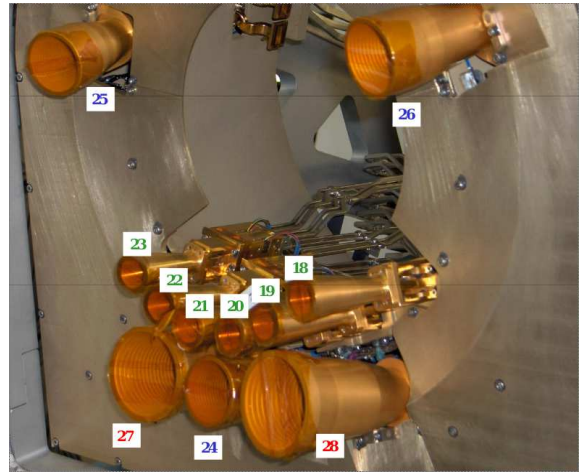}
   		\end{center}
   			\caption{
        Feed horns in the LFI focal plane. Each feed horn is tagged by a label running from LFI18 to LFI28. LFI18 through LFI23 are 70~GHz receivers, LFI24 through LFI26 are 44~GHz receivers and LFI27, LFI28 are 30~GHz receivers.
						}
   \label{fig:LFI_focal_plane}
\end{figure}

The various receivers are tagged with labels from LFI18 to LFI28, as shown in Fig.~\ref{fig:LFI_focal_plane}. Each of the two radiometers connected to the two Ortho Mode Transducer (OMT) arms are labelled as M-0 (``main'' OMT arm) and S-1 (``side'' OMT arm) while the two output detectors from each radiometer are labelled as 0 and 1. Therefore with the label \texttt{LFI18S-1}, for example, we indicate the radiometer \texttt{S} of the receiver LFI18, and with the label \texttt{LFI24M-01} we indicate detector \texttt{1} of radiometer \texttt{M-0} in receiver LFI24.

\section{Comparison of tuning results}
\label{app:EX-SUP}

LNAs Tuning results are presented in the following tables: both bias setting and performance are displayed. As explained in this paper, performance comparison along the different test campaigns (FEM level, RCA, RAA, CSL) is not straight because of the different test conditions, affecting results. Hence, comparison is given only for completeness, since the main objective of the Tuning is to find the optimal bias comparing performance measured in the same test condition, that is within the same test campaign (for a detailed description of the LFI performance see  \cite{2009_LFI_cal_M3} and \cite{2009_LFI_cal_R2})

\begin{table}[h!]
        \begin{center}
        
            \caption{Relative comparison of phase switch optimal biases between different test campaigns. Results are given as a percentage of the CSL $I_1$ and $I_2$ phase switch bias currents.
            }
            \label{tab:PHSW_Tuning_comparison} 
            \begin{tabular}{l c | c c| c c}
              \hline
                        
                &\multicolumn{0}{c}{\textbf{~}}
                &\multicolumn{2}{c}{\textbf{CSL vs. RAA}}
                &\multicolumn{2}{c}{\textbf{CSL vs. RCA}}\\
                  \hline

                \textbf{HORN ID} &\textbf{PH-SW ID} & I1\%& I2\%& $I_1$\%& $I_2$\%  \\
               
		               \hline
                
                    LFI24 & 0(M2)   & 15& 20& 39& 19\\
                    LFI24 & 1(M1)   & 16& 1& 15& 9\\
                    LFI24 & 2(S2)   & 10& 16& 32&14 \\
                    LFI24 & 3(S1)   & 3& 0& 34& 7\\
                   \hline
                    LFI25 & 0(M2)   & 40& 8& 2& 7\\
                    LFI25 & 1(M2)   & 0& 2& 42& 1\\
                    LFI25 & 2(M2)   & 4& 2& 41& 1\\
                    LFI25 & 3(M2)   & 2& 12& 22& 10\\
                   \hline
                    LFI26 & 0(M2)   & 13& 19& 0& 17\\
                    LFI26 & 1(M2)   & 17& 4& 45& 7\\
                    LFI26 & 2(M2)   & 18& 10& 39& 8\\
                    LFI26 & 3(M2)   & 15& 10& 25& 9\\
                   \hline
                    LFI27 & 0(M2)   & 20& 14& 16& 12\\
                    LFI27 & 1(M2)   & 15& 0& 20& 4\\
                    LFI27 & 2(M2)   & 15& 20& 23& 14\\
                    LFI27 & 3(M2)   & 14& 9& 15& 12\\
                    \hline
                    LFI28 & 0(M2)   & 2& 13& 2& 13\\
                    LFI28 & 1(M2)   & 4& 20& 18& 8\\
                    LFI28 & 2(M2)   & 1& 15& 13& 8\\
                    LFI28 & 3(M2)   & 14& 23& 27& 14\\
                    \hline
                 \end{tabular}       
            
        \end{center}
    \end{table}


\begin{table*}[h!]
        \begin{center}
            \caption{Comparison between LNAs optimal bias settings at different Test levels (FEM, RCA, RAA, CSL)  for the 30 GHz and 44 GHz channels. Because bias are measured only at DAE driver level, bias at FEM connectors level are obtained from these by model, considering the cryo-harness resistance of one stand alone RCA or of one entire power group (containing several RCAs)
            }
            
           \label{tab:LNA_Tuning_settings}
            \begin{tabular}{l c| c c c c | c c c c }
                \hline
                        
                &\multicolumn{0}{c}{\textbf{~}}
                &\multicolumn{4}{c}{\textbf{$Vg_1$}}
                &\multicolumn{4}{c}{\textbf{$Vg_2$}}\\
                  \hline

                \textbf{RCA} &\textbf{Ch.} & $FEM$& $RCA$& $RAA$ &$CSL$&$FEM$& $RCA$& $RAA$ &$CSL$\\
              \hline
                
\hline
LFI18	&	0(S2)	&	1.50	&	1.50	&	1.53	&	1.59	&	1.61	&	1.61	&	1.56	&	1.56\\
LFI18	&	1(S1)	&	1.45	&	1.45	&	1.53	&	1.46	&	1.47	&	1.47	&	1.50	&	1.50\\
LFI18	&	2(M1)	&	1.50	&	1.50	&	1.71	&	1.45	&	1.50	&	1.50	&	1.72	&	1.48\\
LFI18	&	3(M2)	&	1.50	&	1.50	&	1.36	&	1.51	&	1.52	&	1.52	&	1.27	&	1.53\\
\hline																			
LFI19	&	0(S2)	&	1.47	&	1.47	&	1.69	&	1.56	&	1.56	&	1.56	&	1.55	&	1.66	\\
LFI19	&	1(S1)	&	1.56	&	1.56	&	1.64	&	1.63	&	1.56	&	1.56	&	1.55	&	1.58	\\
LFI19	&	2(M1)	&	1.50	&	1.50	&	1.61	&	1.59	&	1.52	&	1.52	&	1.48	&	1.54	\\
LFI19	&	3(M2)	&	1.50	&	1.50	&	1.67	&	1.59	&	1.47	&	1.47	&	1.48	&	1.57	\\
\hline																			
LFI20	&	0(S2)	&	1.48	&	1.48	&	1.49	&	1.40	&	1.57	&	1.57	&	1.51	&	1.50	\\
LFI20	&	1(S1)	&	1.48	&	1.48	&	1.58	&	1.49	&	1.57	&	1.57	&	1.59	&	1.66	\\
LFI20	&	2(M1)	&	1.48	&	1.48	&	1.70	&	1.57	&	1.55	&	1.55	&	1.53	&	1.64	\\
LFI20	&	3(M2)	&	1.52	&	1.52	&	1.75	&	1.61	&	1.57	&	1.57	&	1.55	&	1.66	\\
\hline																			
LFI21	&	0(S2)	&	1.49	&	1.49	&	1.48	&	1.59	&	1.61	&	1.61	&	1.57	&	1.64	\\
LFI21	&	1(S1)	&	1.45	&	1.45	&	1.44	&	1.31	&	1.41	&	1.41	&	1.45	&	1.44	\\
LFI21	&	2(M1)	&	1.46	&	1.46	&	1.48	&	1.45	&	1.49	&	1.49	&	1.52	&	1.52	\\
LFI21	&	3(M2)	&	1.45	&	1.45	&	1.55	&	1.43	&	1.41	&	1.41	&	1.37	&	1.44	\\
\hline																			
LFI22	&	0(S2)	&	1.45	&	1.45	&	1.66	&	1.55	&	1.50	&	1.50	&	1.50	&	1.53	\\
LFI22	&	1(S1)	&	1.42	&	1.42	&	1.54	&	1.53	&	1.39	&	1.39	&	1.38	&	1.42	\\
LFI22	&	2(M1)	&	1.42	&	1.42	&	1.34	&	1.53	&	1.42	&	1.42	&	1.54	&	1.45	\\
LFI22	&	3(M2)	&	1.43	&	1.43	&	1.34	&	1.33	&	1.44	&	1.44	&	1.32	&	1.31	\\
\hline																			
LFI23	&	0(S2)	&	1.48	&	1.48	&	1.43	&	1.47	&	1.50	&	1.50	&	1.72	&	1.61	\\
LFI23	&	1(S1)	&	1.53	&	1.53	&	1.52	&	1.40	&	1.53	&	1.53	&	1.28	&	1.63	\\
LFI23	&	2(M1)	&	1.51	&	1.51	&	1.72	&	1.60	&	1.46	&	1.46	&	1.40	&	1.48	\\
LFI23	&	3(M2)	&	1.53	&	1.53	&	1.75	&	1.63	&	1.48	&	1.48	&	1.50	&	1.51	\\
\hline																			
LFI24	&	0(M2)	&	1.20	&	1.33	&	1.39	&	1.38	&	1.20	&	1.18	&	0.85	&	1.05	\\
LFI24	&	1(M1)	&	1.20	&	1.33	&	1.37	&	1.20	&	1.20	&	1.19	&	1.55	&	1.15	\\
LFI24	&	2(S2)	&	1.20	&	1.24	&	1.20	&	1.34	&	1.20	&	1.19	&	1.34	&	1.06	\\
LFI24	&	3(S1)	&	1.21	&	1.05	&	1.20	&	1.20	&	1.21	&	1.19	&	1.06	&	1.19	\\
\hline																			
LFI25	&	0(M2)	&	1.20	&	1.35	&	1.29	&	1.41	&	1.21	&	1.19	&	1.27	&	1.06	\\
LFI25	&	1(M2)	&	1.20	&	1.35	&	1.34	&	1.22	&	1.20	&	1.10	&	1.06	&	1.06	\\
LFI25	&	2(M2)	&	1.20	&	1.25	&	1.39	&	1.34	&	1.20	&	1.19	&	1.15	&	1.15	\\
LFI25	&	3(M2)	&	1.20	&	1.25	&	1.22	&	1.32	&	1.20	&	1.19	&	1.25	&	1.06	\\
\hline																			
LFI26	&	0(M2)	&	1.21	&	1.46	&	1.47	&	1.33	&	1.21	&	1.32	&	1.17	&	1.12	\\
LFI26	&	1(M2)	&	1.20	&	1.46	&	1.48	&	1.48	&	1.20	&	1.10	&	1.22	&	0.94	\\
LFI26	&	2(M2)	&	1.20	&	1.35	&	1.48	&	1.48	&	1.20	&	1.49	&	1.12	&	1.12	\\
LFI26	&	3(M2)	&	1.20	&	1.55	&	1.38	&	1.38	&	1.20	&	0.99	&	1.34	&	1.34	\\
\hline																			
LFI27	&	0(M2)	&	1.40	&	1.69	&	1.70	&	1.69	&	-1.40	&	-1.41	&	-1.40	&	-1.41	\\
LFI27	&	1(M2)	&	1.40	&	1.69	&	1.81	&	1.78	&	-1.40	&	-1.41	&	-1.41	&	-1.83	\\
LFI27	&	2(M2)	&	1.38	&	1.66	&	1.65	&	1.62	&	-1.41	&	-1.78	&	-1.92	&	-1.55	\\
LFI27	&	3(M2)	&	1.41	&	1.67	&	1.93	&	1.83	&	-1.41	&	-1.11	&	-0.98	&	-1.27	\\
\hline																			
LFI28	&	0(M2)	&	1.40	&	1.75	&	1.78	&	1.77	&	-1.40	&	-1.90	&	-1.56	&	-1.56	\\
LFI28	&	1(M2)	&	1.39	&	1.76	&	1.71	&	1.70	&	-1.41	&	-0.93	&	-1.30	&	-1.31	\\
LFI28	&	2(M2)	&	1.40	&	1.56	&	1.59	&	1.70	&	-1.40	&	-1.80	&	-1.86	&	-1.96	\\
LFI28	&	3(M2)	&	1.40	&	1.75	&	1.82	&	1.81	&	-1.41	&	-1.11	&	-1.09	&	-1.09	\\
                     \hline
                      
            \end{tabular}\\
                 \end{center}
    \end{table*}


     \begin{table*}
       \begin{center}
           \caption{Comparison between LNAs Tuning results at different Test levels. Optimal Noise Temperature and Isolation resulting from CSL test campaign are compared with RAA and RCA results. Results at Unit level are not included because of the very different setup conditions: sky and reference horns and loads were replaced by waveguide microwave matched loads and BEMs and DAE were missing.
           }
           \label{tab:LFI_TN_ISO_Tuning_comparison} 

\begin{tabular}{l | c c c | c c c | c  c c | c c c}																										
\hline							
& \multicolumn{6}{c|}{Noise Temperature} &  \multicolumn{6}{c}{Isolation} \\																	
\hline
&\multicolumn{3}{c|}{\textbf{MAIN}}																										
&\multicolumn{3}{c|}{\textbf{SIDE}}
&\multicolumn{3}{c|}{\textbf{MAIN}}																										
&\multicolumn{3}{c}{\textbf{SIDE}} \\
\hline
\textbf{RCA}& $RCA$& $RAA$&$CSL$& $RCA$& $RAA$&$CSL$ & $RCA$& $RAA$&$CSL$& $RCA$& $RAA$&$CSL$  \\							\hline																										
LFI 18	&	36.0	&	N.A.	&	27.7	&	34.5	&	36.6	&	28.1	&	-13.3	&	-12.3	&	-15.4	&	-11.0	&	-12.7	&	-14.6		\\
LFI 19	&	32.3	&	37.2	&	26.4	&	32.9	&	36.4	&	29.6	&	-15.7	&	-13.0	&	-13.8	&	-14.5	&	-20.0	&	-15.8		\\
LFI 20	&	34.7	&	40.8	&	29.1	&	35.9	&	41.8	&	32.3	&	-15.8	&	-20.0	&	-17.3	&	-13.4	&	-12.0	&	-15.7		\\
LFI 21	&	27.9	&	32.4	&	24.0	&	35.4	&	43.2	&	32.6	&	-12.7	&	-20.0	&	-16.3	&	-10.3	&	-7.4	&	-7.8		\\
LFI 22	&	30.6	&	39.1	&	27.8	&	31.1	&	41.6	&	28.6	&	-11.8	&	-11.5	&	N/A	&	-11.8	&	-10.1	&	-14.2		\\
LFI 23	&	35.0	&	39.2	&	30.5	&	32.5	&	51.3	&	31.7	&	-12.2	&	-9.5	&	-16.0	&	-13.8	&	-12.7	&	-19.8		\\
\hline																										
LFI 24	&	15.4	&	N.A.	&	18.1	&	15.8	&	17.3	&	19.1	&	-12.0	&	-20.0	&	18.1	&	-10.4	&	-18.6	&	-19.6		\\
LFI 25	&	17.7	&	18.1	&	17.4	&	18.5	&	17.8	&	17.7	&	-10.7	&	-20.0	&	17.4	&	-11.7	&	-20.0	&	-20.0		\\
LFI 26	&	17.9	&	15.2	&	22.9	&	16.6	&	15.2	&	13.8	&	-11.3	&	-16.3	&	22.9	&	-13.7	&	-20.0	&	-19.2		\\
\hline
LFI 27	&	12.0	&	10.9	&	13.7	&	12.7	&	12.9	&	15.1	&	-12.9	&	-20.0	&	13.7	&	-14.6	&	-20.0	&	-20.0		\\
LFI 28	&	10.5	&	10.7	&	13.8	&	9.9	&	9.7	&	14.4	&	-10.6	&	-16.6	&	13.8	&	-10.4	&	-14.4	&	-13.6		\\
\hline																										

      \end{tabular}
   \end{center}
\end{table*}

\end{document}